\documentclass[twocolumn]{emulateapj}
\usepackage{natbib}
\usepackage{epsfig}
\usepackage{bm}

\providecommand{\adsurl}[1]{\href{#1}{}}

\begin{document}

\newcommand{\clee}{C_{\ell}^{EE}}
\newcommand{\xef}{x_e^{\rm fid}}
\newcommand{\xet}{x_e^{\rm true}}
\newcommand{\dz}{\Delta z}
\newcommand{\zmax}{z_{\rm max}}
\newcommand{\zmin}{z_{\rm min}}
\newcommand{\zmid}{z_{\rm mid}}
\newcommand{\lcdm}{$\Lambda$CDM}
\newcommand{\wmap}{\emph{WMAP}}

\title{Model-independent constraints on reionization from large-scale CMB polarization}
%\shorttitle{Model-independent constraints on reionization}

\author{Michael J. Mortonson$^{1,2}$ and Wayne Hu$^{1,3}$}
\affil{$^{1}$Kavli Institute for Cosmological Physics, 
Enrico Fermi Institute, University of Chicago, Chicago, IL 60637\\
$^{2}$Department of Physics,  University of Chicago, Chicago, IL 60637\\
$^{3}$Department of Astronomy and Astrophysics, University of Chicago, Chicago, IL 60637
}

\begin{abstract}
On large angular scales, the polarization of the CMB contains 
information about the evolution of the average ionization during the epoch
of reionization.   
 Interpretation of the polarization 
spectrum usually requires the assumption of a fixed functional form for the
evolution, e.g.\ instantaneous reionization.
We develop a model-independent method where a small set of
principal components 
 completely encapsulate the 
effects of reionization on the large-angle $E$-mode polarization
for any reionization history 
within an adjustable range in redshift. Using Markov Chain Monte Carlo 
methods, 
we apply this approach to both the 3-year \wmap\ data and 
simulated future data.  \wmap\ data constrain two principal components 
of the reionization history, approximately corresponding to the 
total optical depth and the difference between the contributions to the 
optical depth at high and low redshifts.  
The optical depth is consistent with the constraint found in previous 
analyses of \wmap\ data that assume instantaneous reionization, with 
only slightly larger uncertainty due to the expanded set of models. 
Using the principal component approach, \wmap\ data also place a 
95\% CL upper limit of 0.08 on the contribution to the optical depth from 
redshifts $z>20$. 
With improvements in 
polarization sensitivity and foreground modeling,
approximately five of the principal components can ultimately be measured. 
Constraints on the principal components, which probe the entire reionization 
history, can test models of reionization, 
provide model-independent constraints on the optical depth, and 
 detect signatures of high-redshift reionization.
\end{abstract}


% =====================================================
\section{Introduction}
\label{sec:intro}

The amplitude of the $E$-mode component of 
the cosmic microwave background (CMB)
polarization on large scales provides the current 
best constraint on the Thomson scattering optical depth to 
reionization, $\tau$.  
Using the first three years of data from the \emph{Wilkinson Microwave 
Anisotropy Probe} (\wmap) and making the simple assumption 
that the universe was reionized instantaneously, 
\cite{Speetal06} find $\tau = 0.09 \pm 0.03$. 
Theoretical studies suggest 
that the process of reionization 
was too complex to be well described 
as a sudden transition~\citep[e.g.,][]{BarLoe01}.  
Previous studies have examined how the constraint on $\tau$ depends on 
the evolution of the globally-averaged ionized fraction during 
reionization, $x_e(z)$, for a variety of specific theoretical scenarios. 
If the assumed form of $x_e(z)$ is incorrect, 
the estimated value of $\tau$ can be biased; this bias can be lessened by 
considering a wider variety of reionization histories at the expense of 
increasing the uncertainty in $\tau$~\citep{Kapetal03,Holetal03,Coletal05}.

The angular scales on which CMB polarization from reionization is correlated 
depend on the horizon size at the redshift of the free electrons: 
the higher the redshift, the higher the multipole, $\ell$ 
\citep[e.g.,][]{Zal97,HuWhi97a}. 
Varying $x_e(z)$ changes the relative contributions to the polarization 
coming from different redshifts, and therefore changes the shape of the 
large-scale $E$-mode angular power spectrum, $\clee$. 
Because of this dependence, measurements of the low-$\ell$ $E$-mode 
spectrum should place at least weak constraints on the global reionization 
history in addition to the constraint on the total optical depth. 
Recent studies suggest 
that \wmap\ data provide little information about $x_e(z)$ beyond $\tau$ 
\citep[e.g.,][]{LewWelBat06}, 
but it is worth asking what we can ultimately expect to 
learn about reionization from CMB polarization.

\citet{HuHol03} proposed using a principal component decomposition
of the ionization 
history to quantify the information contained in the large-scale $E$-mode 
polarization.  
The effect of any ionization history on the $E$-modes can be completely 
described by a small number of eigenmode parameters, unlike
a direct discretization of $x_{e}(z)$ in redshift bins.
 Here we extend the methods of \cite{HuHol03} 
using Markov Chain Monte Carlo techniques to find constraints on the 
principal components of $x_e(z)$ using both the 3-year \wmap\ observations 
and simulated future data.

Analytic studies and simulations indicate that reionization is an 
inhomogeneous process, and this inhomogeneity is expected to contribute 
to the small-scale CMB temperature and polarization anisotropies 
\citep[e.g.,][]{Hu00,Ilietal06a,MorHu07,Doretal07}. 
Here we focus on the large-scale $E$-modes only ($\ell \lesssim 100$) 
where such effects can be neglected, so we only consider  
the evolution of the globally-averaged ionized fraction as a function of 
redshift.

In the following section, we describe the principal component method 
for parameterizing the ionization history and show that the effects of 
$x_e(z)$ on large-scale $E$-mode polarization can be encapsulated in a 
small set of parameters. The method allows $x_e(z)$ to be a free 
function of redshift that is not bounded by physical considerations, so 
in \S~\ref{sec:phys} we derive limits that can be placed on the 
model parameters to eliminate most of the unphysical models where 
$x_e>1$ or $x_e<0$. We outline several ways to apply the principal 
component approach to constrain the reionization history with 
large-scale $E$-mode data in \S~\ref{sec:appl}. Using Markov Chain 
Monte Carlo methods, we examine some of these applications in more detail 
in \S~\ref{sec:mcmc} using both 3-year \wmap\ data and simulated 
future CMB polarization data. We summarize our findings and conclude 
in \S~\ref{sec:discuss}.

% =====================================================
\section{Ionization History Eigenmodes}
\label{sec:eigenmodes}

Models with similar total optical depth but different ionization histories
can produce markedly different predictions for the $E$-mode power spectrum.
In an instantaneous reionization scenario, the contribution to the optical 
depth from $x_e(z)$ is concentrated at the lowest redshifts possible 
for a given $\tau$, and the $E$-mode power spectrum for such a model is 
sharply peaked on large scales, at $\ell \lesssim 10$.  The main effect of 
shifting some portion of the reionization history to higher redshifts 
while keeping $\tau$ fixed is to reduce the $E$-mode power on the largest 
scales and increase it on smaller scales. 

\begin{figure}
\centerline{\psfig{file=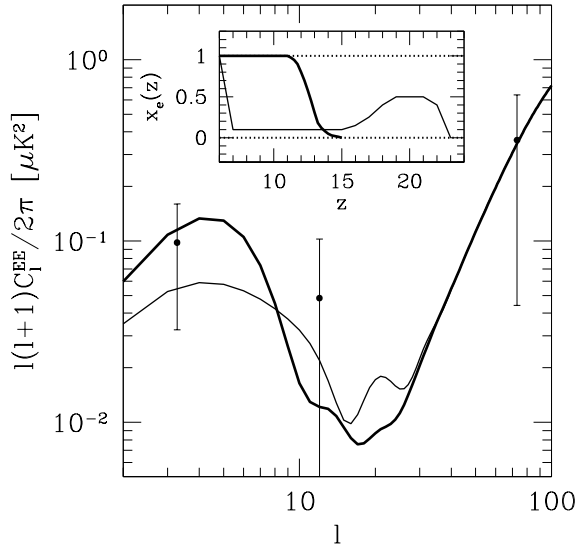, width=3.0in}}
\caption{$E$-mode polarization angular power spectra and 
ionization histories (\emph{inset}) for a nearly-instantaneous 
reionization model with optical depth 
$\tau = 0.105$ (\emph{thick curves}) and an extended, 
``double'' reionization history with $\tau = 0.090$ (\emph{thin}). 
Points with error bars represent the 3-year \wmap\ data from \cite{Pagetal06}.
\vskip 0.25cm}
\label{fig:cl}
\end{figure}

This redistribution of power is illustrated in Figure~\ref{fig:cl} by the ionization histories 
and $\clee$ for two models, one with nearly instantaneous reionization 
and the other with comparable optical depth but with $x_e(z)$ 
concentrated at higher $z$.  In general, a flatter large-scale $E$-mode 
spectrum with power extending out to $\ell \sim 10$-20 is a sign of 
a large ionized fraction at high redshift.  However, there is not 
a one-to-one correspondence between $\ell$ and $z$; the ionized fraction 
within any particular narrow redshift bin affects $\clee$ over a 
wide range of angular scales.
Ionized fractions in adjacent redshift bins have highly correlated 
effects on $\clee$, which makes it difficult to extract from $E$-mode 
polarization data a constraint on $x_e$ at a specific redshift.

% ****************************************
% ****************************************

As suggested by \cite{HuHol03}, one can use principal components of 
the reionization history as the model parameters instead of $x_e$ in 
redshift bins. These components are defined to have uncorrelated 
contributions to the $E$-mode power; since each has a unique effect on 
$\clee$, the amplitudes of the components can be inferred from 
measurements of the large-scale power spectrum. Principal component analysis 
of $x_e(z)$ also indicates which components can be determined best from 
the data. In this section, we describe the principal component method and 
introduce the notation that we will use throughout the paper.

Consider a binned ionization history $x_e(z_i)$, $i\in \{1,2,\ldots,N_z\}$, 
with redshift bins of width $\dz$ spanning $\zmin \leq z \leq \zmax$, where 
$z_1=z_{\rm min}+\dz$ and $z_{N_z}=z_{\rm max}-\dz$ so that 
$N_z+1=(z_{\rm max}-z_{\rm min})/\Delta z$. 
Throughout this paper we assume that the ionized fraction is 
$x_e\approx 1$ for redshifts $z\leq z_{\rm min}$ and $x_e\approx 0$ 
at $z\geq z_{\rm max}$. 
We take $z_{\rm min}=6$, consistent with observations of 
quasar spectra~\citep{FanCarKea06}. 

The principal components of $x_e(z_i)$ are eigenfunctions 
of the Fisher matrix, computed by taking derivatives of 
 $\clee$ with respect to $x_e(z_i)$:
\begin{equation}
F_{ij}=\sum_{\ell=2}^{\ell_{\rm max}}\left(\ell+\frac{1}{2}\right)
 \frac{\partial \ln \clee}{\partial x_e(z_i)}
 \frac{\partial \ln \clee}{\partial x_e(z_j)} \,,
\label{eq:fisher1}
\end{equation}
assuming full sky coverage and neglecting noise.
Since $x_e(z_i)$ only significantly contributes to 
the $E$-mode spectrum at small $\ell$, 
we typically truncate the sum in equation~(\ref{eq:fisher1}) 
at $\ell_{\rm max}=100$
where $\clee$ is dominated by the first acoustic peak.
The derivatives are evaluated 
at a fiducial reionization history,
$\xef(z_i)$. Following \cite{HuHol03} we typically choose 
$\xef(z_i)$ to be constant during reionization, although other 
functional forms  may be used.

Since the effects on $\clee$ of $x_e(z_i)$ in 
 adjacent redshift bins are highly 
correlated, the Fisher matrix contains large off-diagonal elements. 
The principal components $S_{\mu}(z_i)$ are the eigenfunctions of $F_{ij}$,
\begin{equation}
F_{ij}=(N_z+1)^{-2}\sum_{\mu=1}^{N_z}
 S_{\mu}(z_i)\sigma^{-2}_{\mu}S_{\mu}(z_j),
\label{eq:fisher2}
\end{equation}
where the factor $(N_z+1)^{-2}$ is included so that the eigenfunctions and 
their amplitudes have certain convenient properties. The inverse eigenvalues, 
$\sigma^2_{\mu}$, give the estimated variance 
of each principal component eigenmode from the measurement of low-$\ell$ 
$E$-modes. We order the modes so that the best-constrained 
principal components (smallest $\sigma^2_{\mu}$) have the lowest values of 
$\mu$, starting at $\mu=1$.
The noise level and other characteristics of an experiment 
can be included in the construction of the eigenfunctions, but since the 
effect on $S_{\mu}(z)$ is small we always use the noise-free 
eigenfunctions here.

The eigenfunctions satisfy the orthogonality and completeness
relations
\begin{eqnarray}
\int _{\zmin}^{\zmax} dz
~S_{\mu}(z)S_{\nu}(z)&=&(z_{\rm max}-z_{\rm min})\delta_{\mu\nu}\,,
\label{eq:orthog1}\\ 
\sum_{\mu=1}^{N_z} S_{\mu}(z_i)S_{\mu}(z_j)&=& (N_z+1) \delta_{ij}\,.
\label{eq:orthog2}
\end{eqnarray}
The normalization 
of $S_{\mu}(z)$ is chosen so that the eigenfunctions 
are independent of bin width as $\dz\rightarrow 0$.
In equation (\ref{eq:orthog1})
and elsewhere in this paper where there are sums over redshift, 
we assume the continuous limit, 
replacing $\sum_i \Delta z$ by $\int dz$.
As long as the bin width 
is chosen to be sufficiently small, the final results we obtain are 
independent of the redshift binning. We adopt $\Delta z=0.25$ as the 
default bin width.

\begin{figure}
\centerline{\psfig{file=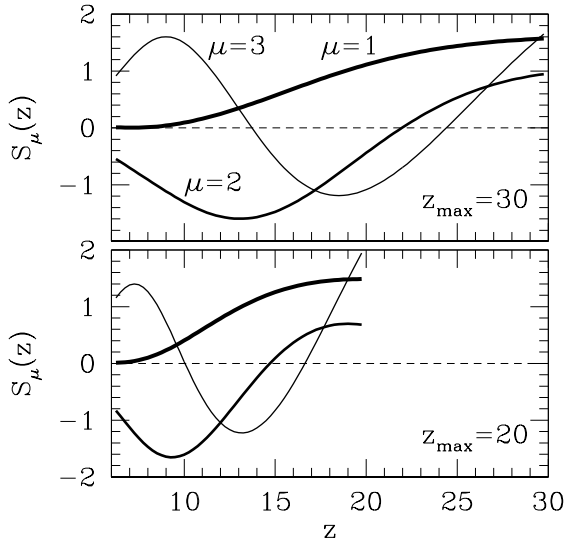, width=3.0in}}
\caption{Eigenfunctions for fiducial models with $\zmax=30$ and constant 
$x_e=0.15$ (\emph{top}), and $\zmax=20$ and $x_e=0.3$ (\emph{bottom}). 
In each case, the first three modes are shown: $\mu=1$ (\emph{thick}), 
2 (\emph{medium}) and 3 (\emph{thin}). For both fiducial models, $\zmin=6$ 
and $\Delta z=0.25$.
\vskip 0.25cm}
\label{fig:eigfn}
\end{figure}

The three lowest-variance eigenfunctions for two different fiducial models 
are shown in Figure~\ref{fig:eigfn}. The lowest eigenmode ($\mu=1$) is an 
average of the ionized fraction over the entire redshift range, weighted at 
high $z$.  The $\mu=2$ mode can be thought of as a difference between 
the amount of ionization at high $z$ and at low $z$, and higher modes 
follow this pattern with weighted averages of $x_e(z)$ that oscillate with 
higher and higher frequency in redshift. Eigenfunctions of fiducial models 
with different values of $\zmax$ have similar shapes with the redshift 
axis rescaled according to the width of $(\zmax-\zmin)$.
The eigenfunctions are mostly insensitive to the choice of ionized 
fraction in the constant-$x_e$ fiducial histories.

% ****************************************
% ****************************************

An arbitrary reionization history can be represented in terms of the 
eigenfunctions as
\begin{equation}
x_e(z)=\xef(z)+\sum_{\mu}m_{\mu}S_{\mu}(z),
\label{eq:mmutoxe}
\end{equation}
where the amplitude of eigenmode $\mu$ for a perturbation 
$\delta x_e(z)\equiv x_e(z)-\xef(z)$ is
\begin{equation}
m_{\mu}=\frac{1}{\zmax-\zmin} 
  \int _{\zmin}^{\zmax} dz~S_{\mu}(z)\delta x_e(z).
\label{eq:xetommu}
\end{equation}
[Note that our conventions for the normalization of 
$m_{\mu}$ and $S_{\mu}(z)$ differ from those of \cite{HuHol03} by factors 
of $(N_z+1)^{1/2}$.] 
Any global ionization history $x_e(z)$ over the range $z_{\rm min}<z<\zmax$ 
is completely specified by a set of mode amplitudes $m_{\mu}$.

If perturbations to the fiducial history are small, $\delta x_e(z) \ll 1$, 
then the mode amplitudes are uncorrelated, with covariance matrix 
$\langle m_{\mu}m_{\nu}\rangle=\sigma_{\mu}^2 \delta_{\mu\nu}$. 
For a fixed fiducial model, however, arbitrary reionization histories 
 generally have $\delta x_e\sim 1$, in which case the amplitudes 
of different modes can become correlated as we discuss in \S~\ref{sec:fisher}.

\begin{figure}
\centerline{\psfig{file=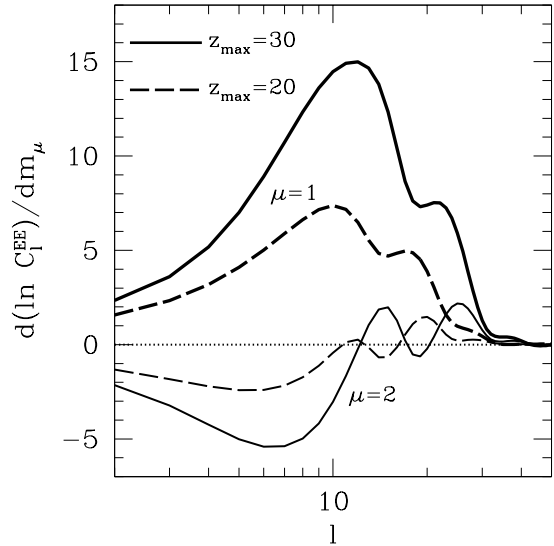, width=3.0in}}
\caption{Change in $\ln \clee$ per unit $m_{\mu}$
for fiducial $x_e(z)$ with $\zmax=30$ (\emph{solid}) and 20 (\emph{dashed}), 
showing the effect on $\clee$ of the first two principal components, 
$\mu=1$ (\emph{thick}) and $2$ (\emph{thin}). 
\vskip 0.25cm
}
\label{fig:zmaxa}
\end{figure}

The main advantage of using the principal component eigenmodes of $x_e(z)$ 
instead of some other parameterization is 
that most of the information relevant for large-scale $E$-modes is 
contained in the first few modes. This means that if one constructs $x_e(z)$ 
from equation~(\ref{eq:mmutoxe}) keeping only the first few terms in the sum 
over $\mu$, then the $E$-mode spectrum of the resulting ionization history 
will closely match that of $x_e(z)$ with all modes included in the sum. 
The effect of each eigenmode on $\clee$ becomes smaller as $\mu$ increases, 
as shown in Figure~\ref{fig:zmaxa} for the first two modes. 
\cite{HuHol03} demonstrated that for a specific fiducial ionization 
history $\xef(z)$ and assumed true history, 
only the first three modes of $x_e(z)$ are needed 
to produce $\clee$ indistinguishable from the true $E$-mode spectrum. 

% ****************************************
% ****************************************

\begin{figure}
\centerline{\psfig{file=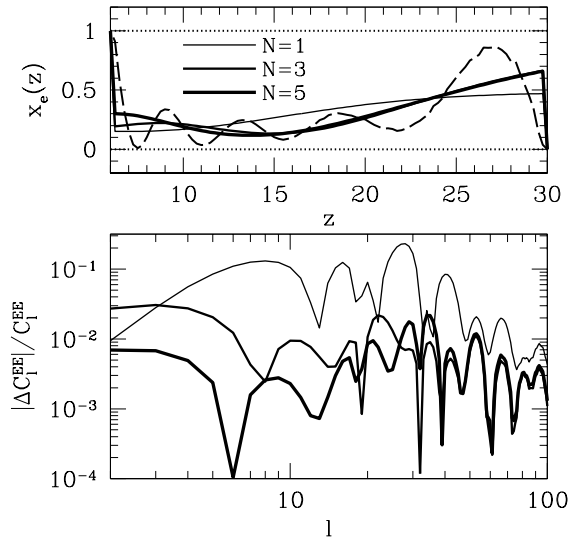, width=3.0in}}
\caption{Test of completeness in $\clee$ for ionization histories with 
truncated sets of eigenfunctions. The base history (\emph{top panel, dashed}) 
is the sum of the first 10 eigenfunctions 
with comparable values of $|m_{\mu}|$ for 
each mode. Solid curves show $x_e(z)$ (\emph{top}) and the error in $\clee$ 
(\emph{bottom}) for histories that retain the first one (\emph{thin curves}), 
three (\emph{medium}), and five (\emph{thick}) eigenmodes. 
The fiducial model used for the eigenmode decomposition has constant 
$x_e=0.15$ from $\zmin=6$ up to $\zmax=30$.
\vskip 0.25cm}
\label{fig:clcomp}
\end{figure}

The ionization histories and corresponding $E$-mode spectra in 
Figure~\ref{fig:clcomp} demonstrate this completeness for a fairly 
extreme model in which the first ten eigenmodes all have 
significant amplitudes.  Even in the simplest case where a single eigenmode 
is used in place of the original $x_e(z)$, the error in $\clee$ is 
only $\sim 10\%$.  With 3-5 modes, the error is a few percent or less at
all multipoles and safely smaller than the cosmic variance of 
\begin{equation}
{\Delta \clee \over \clee} = \sqrt{2 \over 2 \ell +1} \,.
\end{equation}
The top panel of Figure~\ref{fig:clcomp} shows that this completeness 
does not extend to the ionization history itself: $x_e(z)$ constructed from 
as many as five eigenmodes is a poor approximation to the full 
reionization history.

% ****************************************
% ****************************************

From this and other similar tests on the completeness in $\clee$ 
of the lowest-variance eigenmodes we conclude that the first 3-5 modes contain 
essentially all of the information about the reionization history that 
is relevant for  large-scale $E$-mode polarization.
This fact is particularly useful for constraining the global 
reionization history with CMB polarization data using 
Markov Chain Monte Carlo techniques. Since the number of parameters that 
must be added to a Monte Carlo chain  
to describe an arbitrary $x_e(z)$ is relatively small, we can obtain 
constraints from the data that are independent of assumptions about 
the reionization history with minimal added computational 
expense~\citep{HuHol03}.
The exact number of modes required varies depending on the true 
ionization history and the fiducial history, so when analyzing data it is 
a good idea to check that the results do not change significantly when 
the next modes are included in the sum in equation~(\ref{eq:mmutoxe}).

\begin{figure}
\centerline{\psfig{file=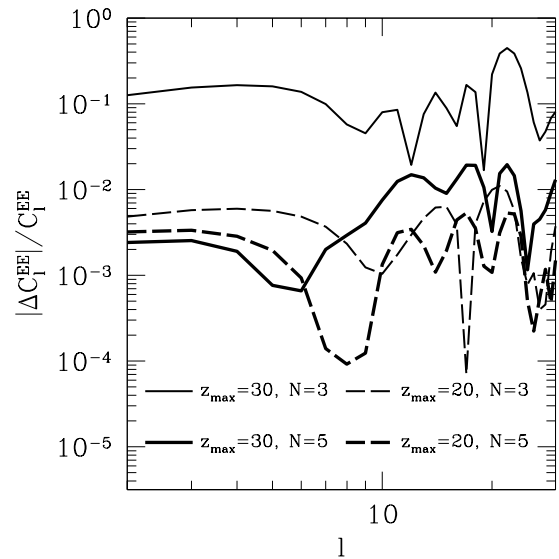, width=3.0in}}
\caption{Error in $\clee$ from $x_e(z)$ constructed 
from the first $N$ eigenmodes of fiducial models with $\zmax=30$ 
(\emph{solid curves}) and 20 (\emph{dashed}), taking $N=3$ (\emph{thin}) 
and $N=5$ (\emph{thick}). The true $x_e(z)$ and $\clee$ 
are assumed to be those of an instantaneous reionization model with $\tau=0.09$.
\vskip 0.25cm}
\label{fig:zmaxb}
\end{figure}

The main caveat to this model independence is that in practice we must 
set some maximum redshift for histories in any particular chain of 
Monte Carlo samples, ignoring any 
contribution to the observed low-$\ell$ $\clee$ from ionization at $z>\zmax$. 
Since the eigenfunctions of the ionization history are stretched in redshift 
as $\zmax$ increases (Fig.~\ref{fig:eigfn}), 
at higher $\zmax$ more modes are needed to accurately 
represent any particular feature in $x_e(z)$. For example, take the true 
ionization history to be instantaneous reionization at $z=11.5$. 
Figure~\ref{fig:zmaxb} shows the error in $\clee$ if we truncate the 
eigenmode sum of equation~(\ref{eq:mmutoxe}) 
at $N$ modes using fiducial histories 
with $\zmax=30$ and $\zmax=20$. For each fiducial model, the error decreases 
as the number of modes in the sum increases, but the error at fixed $N$ is 
larger for $\zmax=30$ than $\zmax=20$. 
The requirement of retaining a larger set of parameters as $\zmax$ 
increases makes it less practical to study models with significant 
reionization at extremely high redshifts 
[$\zmax \gtrsim 100$; e.g., \cite{NasChi04,KasKawSug04}], but even for 
$\zmax$ as high as $\sim 40$ the number of eigenmodes needed 
is reasonably small ($N\lesssim 5$).

% ****************************************
% ****************************************

While theories of reionization provide useful priors on $\zmax$ in 
the context of specific models, it would be better to be able to constrain 
$\zmax$ empirically by measuring the $E$-mode power accurately up 
to $\ell \sim 100$. 
The current 3-year \wmap\ polarization data have 
high enough signal-to-noise to be useful for parameter constraints only at 
$\ell < 24$ \citep{Pagetal06}, so their sensitivity to 
high-$z$ reionization is limited.  
Given that ionization at some redshift generates polarization out
to a maximum $\ell$, 
 \wmap\ data can still place weak bounds on the total optical depth 
contribution above a certain redshift, even if it cannot distinguish whether
these contributions arise from redshifts above a chosen $\zmax$
as we discuss in \S~\ref{sec:wmap}.
The values we choose here for $\zmax$ are 
partly influenced by the fact that few theoretical 
reionization scenarios predict an ionized 
fraction at $z\gtrsim 30$ that would significantly affect large-scale $\clee$. 
Future data should better constrain $\clee$ at higher multipoles, 
allowing useful limits to be placed on $\zmax$ from the data alone.

The need to consider a limited range in redshift is shared by other 
methods, for example those that constrain binned $x_e(z_i)$ instead of 
the eigenmode amplitudes \citep{LewWelBat06}. As already mentioned, the 
principal component approach has the unique advantages that results can be
made independent of bin width and that relatively few extra parameters 
are required.  However, this approach also has a unique difficulty in that 
the physicality of the reionization history 
[$0\leq x_e(z)\leq 1$] is not built in to the method.   Hence constraints derived
from measurements of the eigenmodes can be weaker than those from 
a method that enforces physicality.
We can, however, place some prior constraints on the set of eigenmode 
amplitudes that must be satisfied by any physical model, as we discuss in 
the next section.

% =====================================================
\section{Priors from Physicality}
\label{sec:phys}

Although the actual ionized fraction must be between 0 and 1 (neglecting 
helium reionization  and the small residual ionized fraction after 
recombination), there is nothing in the construction of $x_e(z)$ from 
the eigenmodes in equation~(\ref{eq:mmutoxe}) that ensures that 
the ionized fraction will 
obey these limits. Whether or not $x_e$ has a physical value at a 
particular redshift depends on the amplitudes of all of the 
principal components. 
Even for a physical ionization history, 
 the truncated sum up to mode $N$, 
\begin{equation}
x_e^{(N)}(z)\equiv \xef(z)+\sum_{\mu=1}^N m_{\mu} S_{\mu}(z),
\label{eq:trunc}
\end{equation}
is not necessarily bounded by 0 and 1 at all redshifts. 
While formally it is possible to evaluate $\clee$ for reionization 
histories with unphysical values of $x_e$, we would like to 
eliminate as much as possible those models for which 
the full sum of the eigenmodes, 
$x_e(z)=\lim_{N\to\infty}x_e^{(N)}(z)$, is unphysical.

We find the largest and smallest values of 
$m_{\mu}$ that are consistent with $x_e(z)\in [0,1]$ for all $z$ using 
the definition of $m_{\mu}$ in equation~(\ref{eq:xetommu}). 
We are free to choose $\xef(z)\in [0,1]$ so that  
 $-\xef(z)\leq\delta x_e(z)\leq 1-\xef(z)$, where the lower limit 
is strictly negative or zero and the upper limit is positive or zero. 
For a particular mode $\mu$, 
the choice of $\delta x_e(z)$ that maximizes $m_{\mu}$ is
\begin{equation}
\delta x_e^{\rm max}(z)=\left\{ \begin{array}{rl}
-\xef(z), & S_{\mu}(z)\leq 0 \\
1-\xef(z), & S_{\mu}(z)>0.
\end{array}\right.
\label{eq:dxemax}
\end{equation}
Using this in equation~(\ref{eq:xetommu}) gives an upper limit on $m_{\mu}$. 
Similarly, a lower limit can be obtained by reversing the signs of the 
inequalities in equation~(\ref{eq:dxemax}).  
The resulting physicality bounds  
are $m_{\mu}^{(-)}\leq m_{\mu}\leq m_{\mu}^{(+)}$, where
\begin{equation}
m_{\mu}^{(\pm)} = \int _{\zmin}^{\zmax} dz 
\frac{S_{\mu}(z)[1-2\xef(z)]\pm |S_{\mu}(z)|}{2(\zmax-\zmin)}.
\label{eq:physbounds}
\end{equation}
If $m_{\mu}$ violates these bounds for any $\mu$, the reionization history 
 is guaranteed to be unphysical for some 
range in redshift. The opposite is not true, however: even if all $m_{\mu}$ 
satisfy equation~(\ref{eq:physbounds}), $x_e(z)$ may still be unphysical 
for some $z$.

The parameter space that physical models may occupy is restricted 
further by an inequality that must be satisfied by all eigenmodes 
simultaneously. Assume for simplicity that the fiducial model has 
a constant ionized fraction, $\xef\in [0,1]$, for $\zmin<z<\zmax$. 
Any physical reionization history $x_e(z)$ must satisfy
\begin{equation}
\int _{\zmin}^{\zmax} dz [x_e(z)-\xef]^2 \leq (\zmax-\zmin) f,
\label{eq:phys2}
\end{equation}
where $f\equiv \max[(\xef)^2,(1-\xef)^2]$.
Using equation~(\ref{eq:mmutoxe}), the left side of the inequality can also be 
written
\begin{eqnarray}
\int _{\zmin}^{\zmax} dz [x_e(z)-\xef]^2&=&\int _{\zmin}^{\zmax} dz
 \left[\sum_{\mu} m_{\mu} S_{\mu}(z) \right]^2 \nonumber\\
&=& (\zmax-\zmin) \sum_{\mu} m_{\mu}^2 \,,
\label{eq:phys3}
\end{eqnarray}
where the second line follows from the orthogonality of the eigenfunctions 
(eq.~[\ref{eq:orthog1}]). Comparing equations~(\ref{eq:phys2}) 
and~(\ref{eq:phys3}) we obtain a constraint on the sum of the 
squares of the mode amplitudes,
\begin{equation}
\sum_{\mu} m_{\mu}^2 \leq f,
\label{eq:physbounds2}
\end{equation}
where $0.25\leq f\leq 1$, depending on the value of $\xef$. 
As with the physicality bounds of equation~(\ref{eq:physbounds}), 
this 
upper limit is a necessary but not sufficient condition for physicality.

Since in practice we can only constrain a limited set of eigenmodes, 
the uncertainty in modes higher than the first few prevents us from 
simply excluding all models where $x_e<0$ or $x_e>1$ at any redshift 
because the higher modes can have a significant effect on $x_e(z)$. 
However, as shown in \S~\ref{sec:eigenmodes}, the higher modes do not 
affect the polarization power spectrum since the 
high-frequency oscillations in 
redshift of higher modes are averaged out.  Similarly, such eigenmodes 
have a small effect on the optical depth from a sufficiently large 
range in redshift. For example, the optical depth from $15<z<30$ due to 
modes $\mu>5$ subject to the physicality constraints of this section 
can be no larger than $\sim 0.01$, and is likely to be smaller 
for realistic reionization scenarios. Because of this, we assume that 
we can place priors on the \emph{optical depth} that correspond to $0 \le x_e \le 1$ 
 over the relevant range in redshift. Reionization histories 
with an unphysical 
optical depth over a large range in redshift are considered to be 
unphysical models since the addition of higher modes can not perturb the 
optical depth enough to give it a physical value.

% =====================================================
\section{Applications of the Principal Component Method}
\label{sec:appl}

Once constraints on the principal components of the reionization history 
have been obtained from CMB polarization data, there are several ways 
to use those constraints to place limits on 
observables such as $\tau$ or to test 
theories of reionization \citep{HuHol03}. 
We describe some possible applications in this section, 
and in \S~\ref{sec:mcmc} we put these ideas into practice using 
the 3-year \wmap\ data and simulated future data.

As mentioned in \S~\ref{sec:intro}, the constraint on the total optical 
depth to reionization depends on the assumed model for $x_e(z)$. 
The principal component method allows us to explore 
all globally-averaged ionization 
histories within a chosen redshift range, $\zmin<z<\zmax$. 
For a given set of eigenmode amplitudes, $\{m_{\mu}\}$, 
equation~(\ref{eq:mmutoxe}) yields the corresponding ionization history 
which can then be integrated to find the optical depth between any two
redshifts $z_{1}$ and $z_{2}$,
\begin{equation}
\tau(z_1,z_2) = 0.0691(1-Y_p)\Omega_b h
\int_{z_1}^{z_2} dz \frac{(1+z)^2}{H(z)/H_0}x_e(z).
\label{eq:tauz1z2}
\end{equation}
In particular, the total optical depth to reionization is
\begin{equation}
\tau = \tau(0,\zmin)+\tau(\zmin,\zmax),
\label{eq:tau}
\end{equation}
where $\tau(0,\zmin)\approx 0.04$ for $\zmin=6$. 
The principal component approach provides a model-independent way to 
constrain $\tau$, so we expect the results to be unbiased and the uncertainty 
in $\tau$ to accurately reflect the present uncertainty about 
$x_e(z)$.  Moreover, the information about $\tau(\zmin,\zmax)$
 is encapsulated in the first few eigenmodes, i.e. the 
truncated ionization history of equation~(\ref{eq:trunc}), 
$x_e^{(N)}(z)$, with $N=3$-5 for typical fiducial models.

The total optical depth and the first principal component amplitude, $m_1$, 
are similar quantities in that they are both averages of $x_e(z)$ weighted 
at high $z$. As we show in the next section, CMB $E$-mode polarization data 
can constrain higher eigenmodes as well ($\mu \geq 2$). 
In particular, the second mode should be the next best constrained 
quantity since it is constructed to have the smallest variance after $m_1$.
Since $m_2$ is related to the difference in $x_e$ between  
high redshift and low redshift (see Fig.~\ref{fig:eigfn}), 
we might guess that besides total $\tau$ the data are mainly 
constraining the fraction of optical depth coming from 
high redshift versus low redshift.  
Given any set of mode amplitudes defining an ionization history, we can 
compute a low-$z$ optical depth, $\tau(\zmin,\zmid)$, and a high-$z$ 
optical depth, $\tau(\zmid,\zmax)$, for some choice of intermediate 
redshift, $\zmid$.

Note that if either of the redshift ranges $[\zmin,\zmid]$ or 
$[\zmid,\zmax]$ is too narrow, constraints on the partial optical depths  
will be influenced by the physicality priors 
(since $0\leq x_e\leq 1$ sets limits 
on $\tau$ in redshift intervals as discussed in \S~\ref{sec:phys}) 
and by the uncertainty in modes higher than those included in the chain, 
which have greater effect on the optical depth in narrower redshift 
intervals. For these reasons, $\zmid$ should be chosen to be not too close 
to either $\zmin$ or $\zmax$.

Given an appropriately chosen value of $\zmid$, constraints on the 
principal components can be converted into constraints on the  
optical depths at high and low redshift. These partial optical depths 
are observables in the sense that high-$z$ and low-$z$ 
optical depth affect the 
large-scale $E$-modes over different ranges of multipoles.
For example, compare the two reionization models 
in Figure~\ref{fig:cl}. Both have similar total optical depth, but in 
one case $\tau$ only comes from $z<15$, resulting in more power at 
$\ell < 8$ and less power at $\ell > 8$ than the other model in which 
the optical depth primarily comes from $z>15$.

Besides learning about the relative amount of ionization at $z<\zmid$ 
and $z>\zmid$, the partial optical depth constraints also provide a 
way to empirically set the maximum redshift, $\zmax$. 
If some set of data are found to place a tight upper bound 
on $\tau(\zmid,\zmax)$ 
for fairly conservative (i.e. large) values of $\zmid$ and $\zmax$, then 
for analyses of future data this value of $\zmid$ can be used as a new, lower 
value for $\zmax$ since 
optical depth from higher redshifts is small. 
This approach assumes that there is 
essentially no reionization earlier than the original $\zmax$, but as long 
as this initial maximum redshift is taken to be large the presence or 
absence of high-$z$ ionization can be tested in this way over a wide 
range of redshifts. We explore this idea further in \S~\ref{sec:mcmc}.

While we do not examine constraints on specific reionization models in this 
paper, the principal component approach is well suited to model testing. 
Consider a model of the global reionization history, $x_e(z;\bm{\theta})$, 
parameterized by $\bm{\theta}$. This could be a simple toy model (for example, 
instantaneous reionization where $\bm{\theta}$ is a single parameter, 
the redshift of reionization) or a more physical model where $\bm{\theta}$ 
might include parameters that govern properties of the ionizing sources. 
For a particular choice of $\xef(z)$, the principal 
component amplitudes of the reionization model follow from 
equation~(\ref{eq:xetommu}), giving a set of mode amplitudes that 
depend on the model parameters, $\{m_{\mu}(\bm{\theta})\}$. 
Then constraints on the mode amplitudes from CMB polarization data 
(defined using the same fiducial history) can be mapped to 
constraints on the parameters of the reionization model. 

Applying this method to a model that has as one of its parameters a 
maximum redshift allows one to obtain constraints on $\zmax$ within a 
class of theoretical models.
Although obtaining good constraints on model parameters does not necessarily
imply validation of the model class, this procedure  provides a 
straightforward method by which different model 
classes can tested and falsified within a single analysis.

Generating Monte Carlo chains to find constraints on 
$\{m_{\mu}\}$ as described in the next section can be a somewhat 
time-consuming process, but it only needs to be done once 
per redshift range, after which any model of the global 
reionization history within this range $\zmin<z<\zmax$
can be tested using the same parameter chains. 
In cases where constraints from the data turn out 
to be close to Gaussian, the covariance matrix of the principal components 
can be used in place of the full Monte Carlo chains, reducing the 
amount of information needed for model testing from $\sim 10^5$ numbers 
in chains of Monte Carlo samples to only 
$N(N+3)/2$ numbers when $N$ eigenmodes are included in the chains 
[$N$ mean values plus $N(N+1)/2$ entries in the covariance matrix, 
$\langle m_{\mu}m_{\nu}\rangle$].
However, near-Gaussian constraints on $\{m_{\mu}\}$ are likely to be 
possible only for certain realizations of future data, as discussed in 
the next section.

% =====================================================
\section{Markov Chain Monte Carlo Constraints on Eigenmodes}
\label{sec:mcmc}

We find reionization constraints from CMB polarization data using 
Markov Chain Monte Carlo (MCMC) techniques to explore the principal 
component parameter space \citep[see e.g.][]{Chretal01,KosMilJim02,Dunetal04}. 
Chains of Monte Carlo samples are generated using the publicly available 
code CosmoMC\footnote{http://cosmologist.info/cosmomc/} \citep{LewBri02}, 
which includes the code CAMB \citep{Lewetal00} for computing the 
theoretical angular power spectrum at each point in the $\{m_1,\ldots,m_N\}$ 
parameter space. We have modified both codes to allow 
specification of an arbitrary reionization history calculated from a set 
of mode amplitudes using equation~(\ref{eq:mmutoxe}). 

The principal component amplitudes are the only parameters 
allowed to vary in the chains.
Since nearly all the information in $\clee$ from 
reionization is contained within the first few eigenmodes of $x_e(z)$, 
we include the modes $\{m_1,\ldots,m_N\}$ in each chain with $3\leq N\leq 5$. 
We use only the $E$-mode polarization data for parameter constraints, 
and assume that the values of the 
standard \lcdm\ parameters (besides $\tau$) are fixed by measurements of the 
CMB temperature anisotropies. 
This leads to a slight underestimate of the error on $\tau$ as we discuss 
later, but to a good approximation the effect of $x_e(z)$ on the large-scale 
$E$-modes is independent of the other parameters. 

Specifically, we take  
$\Omega_bh^2=0.0222$, $\Omega_ch^2=0.106$, $100\theta=1.04$ (corresponding 
to $h=0.73$), $A_s e^{-2\tau}=1.7 \times 10^{-9}$, and $n_s=0.96$, 
consistent with results from the most recent version of the \wmap\ 3-year 
likelihood code. When computing the optical depth to reionization we 
take the primordial helium fraction to be $Y_p=0.24$. 
We also assume that $\zmin=6$, so that the optical depth contributed at 
lower redshifts is fixed at $\tau(\zmin)\approx 0.04$. 
The remaining total optical depth from reionization,
$\tau(\zmin,\zmax)$, is determined by the 
values of $\{m_{\mu}\}$ for each sample in the chains.
The default bin width for our 
fiducial models $\xef(z)$ is $\dz=0.25$, which is small enough that 
numerical effects related to binning should be negligible. 

To get accurate results from MCMC analysis, 
it is important to make sure that the 
parameter chains contain enough independent samples 
covering a sufficient volume of parameter space so that the density of 
the samples converges to the actual posterior 
probability distribution. 
For each scenario that we study, we run 4 separate chains 
until the Gelman and Rubin convergence statistic $R$, 
corresponding to the ratio of the variance of parameters between chains to 
the variance within each chain, satisfies 
$R-1<0.01$ \citep{GelRub92,BroGel98}. The convergence diagnostic of 
\cite{RafLew92} is used to determine how much each chain must be thinned 
to obtain independent samples.  Both of these statistics and other 
diagnostic measures are computed automatically by CosmoMC.

Since $x_e(z)$ during reionization must match onto $x_e\approx 1$ at 
$z=z_{\rm min}$ and $x_e\approx 0$ at $z=\zmax$, there are often sharp 
transitions at these redshifts since nothing in 
equation~(\ref{eq:mmutoxe}) forces $x_e(z)$ to satisfy these 
boundary conditions. 
To avoid problems with the time integration in CAMB, we 
smooth the reionization history by convolving $x_e(z)$ with a Gaussian of 
width $\sigma_z=0.5$.

As described in previous sections, we assume that $0\leq x_e\leq 1$ 
between $\zmin$ and $\zmax$. This assumption neglects helium reionization, 
which can make $x_e$ slightly larger than unity, and the small 
residual ionized fraction remaining after recombination that 
prevents $x_e$ from ever being exactly zero. 
These are relatively small effects, 
especially since we are not placing constraints on $x_e(z)$ directly but 
rather on weighted averages of $x_e$ over redshift. The residual 
ionized fraction at $z>\zmax$ is accounted for in the Monte 
Carlo exploration of reionization histories.

In \S~\ref{sec:wmap}, we examine the current 
constraints from the 3-year \wmap\ data. 
We then provide forecasts for principal component constraints with idealized, 
cosmic variance-limited, simulated data in \S~\ref{sec:cvdata}.
In each case the likelihood 
computation includes only the $E$-mode polarization data, up to $\ell=100$ for 
simulated data and $\ell=23$ for \wmap ; the likelihood code for \wmap\ 
does not use $\clee$ at smaller scales due to low signal-to-noise.
At multipoles $\ell \gtrsim 100$ the global reionization history only affects 
the amplitude of the angular power spectra, which we fix by setting 
$A_s e^{-2\tau}$ constant in the Monte Carlo chains.
A comparison of the MCMC results with the Fisher matrix 
approximation follows in \S~\ref{sec:fisher}.

% --------------------------------------------
\subsection{Constraints from \wmap}
\label{sec:wmap}

In our analysis of \wmap\ data we use the 3-year \wmap\ likelihood code, 
with settings chosen so that likelihoods include only contributions from 
the low-$\ell$ $E$-mode polarization.  In this regime, the code 
computes model likelihoods with 
a pixel-based method instead of using the angular power 
spectrum~\citep{Pagetal06}.
Since the maximum redshift 
at which there is still a significant ionized fraction is uncertain, we 
generate Monte Carlo chains using different fiducial models with 
$10\leq \zmax \leq 40$. 
To avoid possible bias due to neglecting the possibility of high-redshift 
reionization, we focus here on the results obtained using 
the more conservative values of $\zmax=30$ and 40.

\begin{figure*}
\centerline{\psfig{file=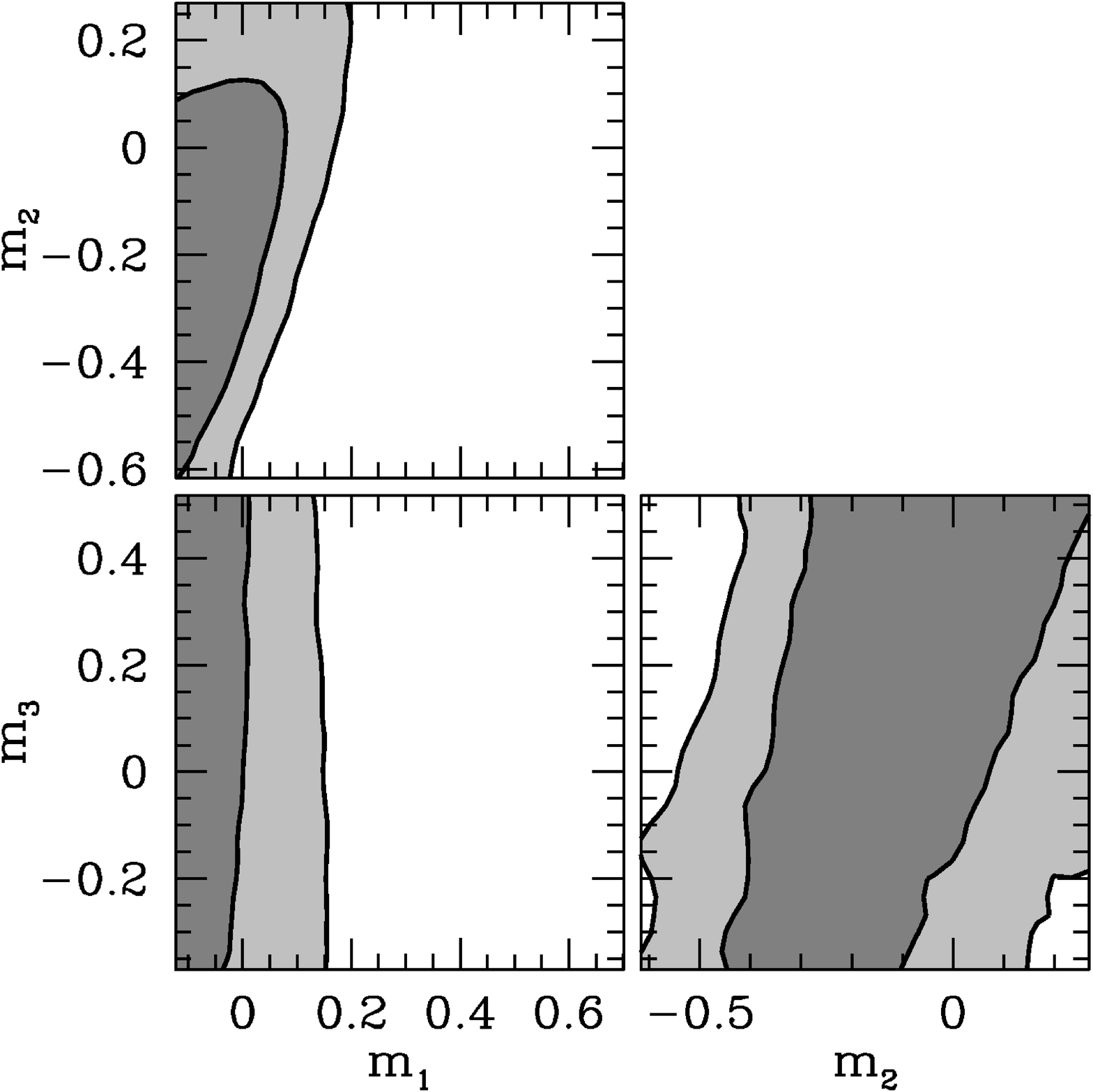, width=3.0in}
\psfig{file=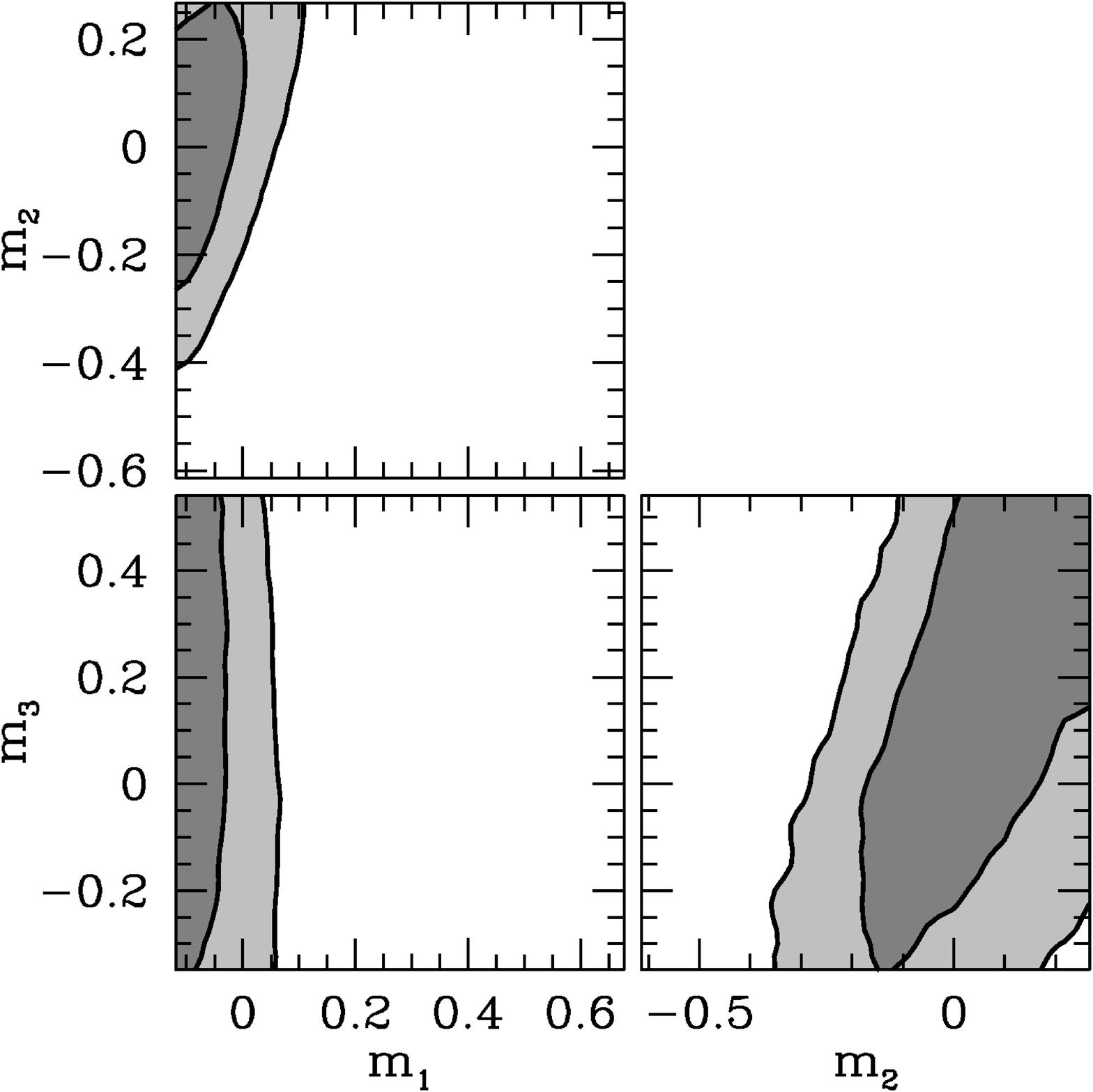, width=3.0in}}
\caption{Marginalized 2D contours 
(\emph{dark shading}: 68\% CL; \emph{light}: 95\%)  for
the first three principal components of $x_e(z)$ from MCMC analysis of 
3-year \wmap\ $E$-mode polarization data, 
using fiducial reionization histories with $\zmax=30$ (\emph{left}) 
and $\zmax=40$ (\emph{right}). Note that the mode amplitudes $m_{\mu}$ 
are defined differently in the left and right panels since the 
fiducial models are different.
\vskip 0.25cm
}
\label{fig:mcmcwmap}
\end{figure*}

The MCMC constraints on the first three principal components from \wmap\ are 
shown in Figure~\ref{fig:mcmcwmap} using fiducial models with $\zmax=30$ and 
$\zmax=40$. The marginalized constraints are plotted within the physicality 
bounds of equation~(\ref{eq:physbounds}), 
$m_{\mu}^{(-)}\leq m_{\mu}\leq m_{\mu}^{(+)}$, so the size of the 
contours inside each box gives an idea of the constraining power of the 
data within the space of potentially physical models. 
For both fiducial histories, the data 
 place a strong upper limit on $m_1$ and weakly constrain $m_2$. 
It is important to note that 
although the parameters $\{m_1,m_2,m_3\}$ have the same names in both the
$\zmax=30$ and $\zmax=40$ plots, they are defined with respect to different 
fiducial models and so we do not expect the contours in the left and right 
plots in Figure~\ref{fig:mcmcwmap} to agree exactly. 
The qualitative similarity 
between the contours simply reflects the fact that the eigenfunctions 
for different $\zmax$ have similar shapes (Figure~\ref{fig:eigfn}).

% ****************************************
% ****************************************

Since $m_1$ and $\tau$ are both averages of $x_e$ from $\zmin$ to $\zmax$ 
with more weight at high $z$ than low $z$, the strong constraint on 
$m_1$ mostly reflects the ability of the data to constrain the total 
optical depth to reionization. Indeed, $m_1$ and $\tau$ are strongly 
correlated in the Monte Carlo chains, while the correlations between 
$\tau$ and higher principal components of $x_e(z)$ are weaker.

Relative to the physicality bounds, the constraints 
in Figure~\ref{fig:mcmcwmap} are 
stronger for the $\zmax=40$ chains than for $\zmax=30$; this is 
to be expected since the physicality bounds on $\{m_{\mu}\}$ permit 
a greater variety of ionization histories, and in particular a 
wider range in $\tau$, when $\zmax$ is larger.
The range of eigenmode amplitudes allowed by 
the limits of equation~(\ref{eq:physbounds}) 
is nearly independent of $\zmax$, but the effect of a 
unit-amplitude principal component on $\clee$ increases with $\zmax$ 
as illustrated in Figure~\ref{fig:zmaxa}.

As described in \S~\ref{sec:appl}, principal component constraints 
from MCMC yield model-independent constraints on the total optical 
depth. For each Monte Carlo sample of the principal components we compute 
$\tau$ using equation~(\ref{eq:tau}). The constraints on $\tau$ are 
listed in Table~\ref{tab:tau} for 
various values of $\zmax$ and $N$, the number of modes in the Monte Carlo 
chains. 
The uncertainty in $\tau$ from the Monte Carlo chains 
is slightly underestimated because 
we fix all cosmological parameters that are not directly related to 
reionization. To estimate the effect of this assumption, we 
compare constraints on $\tau$ in two cases where instantaneous 
reionization is assumed instead of using the model-independent 
principal component approach: 
one in which the non-reionization parameters are 
fixed as in the rest of our Monte Carlo chains, and the other, from the 
3-year \wmap\ analysis of \cite{Speetal06}, in which these  parameters 
are allowed to vary. These constraints, listed in rows 4 and 5 of 
Table~\ref{tab:tau}, suggest that fixing \lcdm\ parameters besides $\tau$ 
reduces $\sigma_{\tau}$ by $\sim 10\%$.

\begin{table}
\caption{Constraints on total optical depth to reionization.}
\begin{center}
\begin{tabular}{llcclc}
\hline
\hline
& Use PCs &  &  & Fix other & \\
Data & of $x_e(z)$?\tablenotemark{a} & ~$\zmax$~ & ~$N$~ & parameters?\tablenotemark{b} & $\tau$ \\
\hline
WMAP3 & Yes & 20 & 3 & Yes & $0.098\pm0.025$ \\
WMAP3 & Yes & 30 & 5 & Yes & $0.106\pm0.027$ \\
WMAP3 & Yes & 40 & 5 & Yes & $0.107\pm0.029$ \\
WMAP3 & No  & -- & -- & Yes & $0.096\pm0.027$ \\
WMAP3 & No  & -- & -- & No & $0.089\pm0.030$\tablenotemark{c} \\
\hline
CV\tablenotemark{d} & Yes & 20 & 3 & Yes & $0.103\pm0.004$ \\
CV & Yes & 30 & 5 & Yes & $0.108\pm0.005$ \\
CV & No & -- & -- & Yes & $0.108\pm0.003$ \\
\hline
\tablenotetext{a}{If not using principal components, reionization is assumed to occur instantaneously.}
\tablenotetext{b}{i.e. $\Omega_bh^2$, $\Omega_ch^2$, $h$, $A_s e^{-2\tau}$, and $n_s$. Other non-reionization parameters are always fixed.}
\tablenotetext{c}{From \cite{Speetal06}.}
\tablenotetext{d}{The true history assumed here is instantaneous reionization 
with $\tau=0.105$, using the cosmic variance-limited realization of $\clee$ in the left 
panel of Figure~\ref{fig:clcv}.}
\end{tabular}
\end{center}
\label{tab:tau}
\end{table}

% ****************************************
\begin{figure}
\centerline{\psfig{file=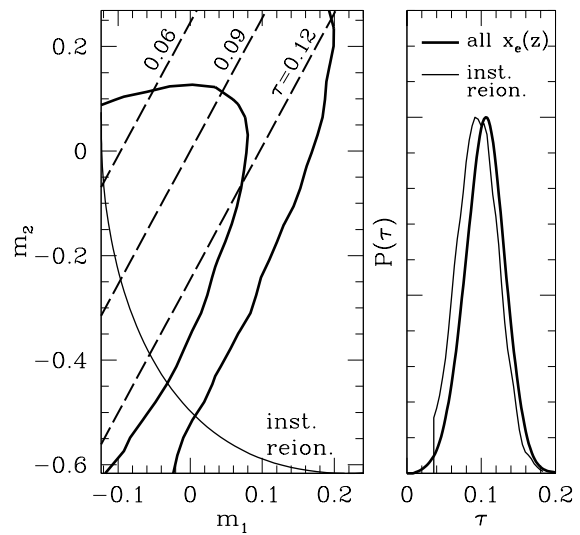, width=3.0in}}
\caption{\emph{Left panel}: 
Marginalized 2D contours (68 and 95\%) from a $\zmax=30$ \wmap\ 
chain of Monte Carlo samples 
(\emph{thick curves}), lines of constant $\tau=\{0.06,0.09,0.12\}$ 
(\emph{dashed lines}), and the set of 
one-parameter instantaneous reionization models 
(\emph{thin curve}) plotted in the $m_1-m_2$ plane. 
Higher eigenmodes ($\mu\geq 3$), which only weakly influence $\tau$, 
are fixed to $m_{\mu}=0$ for the purposes of plotting the 
constant-$\tau$ lines.
\emph{Right panel}: posterior probability of the optical depth from 
3-year \wmap\ data, $P(\tau)$, for 
arbitrary $x_e(z)$ with $\zmin=6$, $\zmax=30$, and $N=5$ principal 
components (\emph{thick curve}), and for the one-parameter 
instantaneous reionization model with a $\zmin=6$ prior (\emph{thin curve}).
\vskip 0.25cm
}
\label{fig:tauinst}
\end{figure}
% ****************************************

In all cases where we fix the other 
cosmological parameters, the uncertainty in $\tau$ is similar 
regardless of whether we use principal components to explore a variety 
of reionization histories or restrict the analysis to 
instantaneous $x_e(z)$. This is somewhat surprising, since in general 
one would expect that expanding the model space for $x_e(z)$ would 
increase the estimated error on $\tau$. 
The physicality priors may be responsible for reducing $\sigma_{\tau}$ 
slightly --- top-hat priors on $\{m_{\mu}\}$ induce 
a prior on $\tau$ that is flat over a certain range 
but falls off approximately linearly at the edges --- but 
even after accounting for priors, 
the optical depth constraint is robust to replacing the 
instantaneous reionization assumption with a model-independent analysis. 

The insensitivity of the constraint on $\tau$ to the set of models 
considered is partly due to the fact that the degeneracy in the 
eigenmode constraints is aligned in the direction of constant $\tau$, 
as shown in Figure~\ref{fig:tauinst} in the $m_1-m_2$ plane. 
The set of instantaneous reionization models, plotted as a curve 
in Figure~\ref{fig:tauinst}, cuts across the \wmap\ constraints on more 
general models in a region of high posterior 
probability where the distribution of samples varies slowly along 
lines of constant $\tau$. This large overlap between the general 
reionization histories favored by the data and the line of 
instantaneous models is the main reason why the probability distributions 
$P(\tau)$, plotted in the right panel of Figure~\ref{fig:tauinst}, 
are similar for the two classes of models. [The sharp cutoff at low $\tau$ 
in the instantaneous reionization $P(\tau)$ comes from our 
$\zmin=6$ prior.] 
Note that the fact that the 
instantaneous reionization curve passes through the middle of 
the 68\% confidence region also indicates that models with rapid 
reionization are not at all disfavored by the 3-year \wmap\ data.
We use two methods to find the posterior distribution 
for $\tau$ in the instantaneous reionization case: one is the usual 
approach of varying the optical depth (or reionization redshift) in a 
Monte Carlo chain, and the other involves computing $P(\tau)$ for 
a subset of samples from chains in which principal components of $x_e(z)$ 
are varied, selecting only those samples with $\{m_{\mu}\}$ values close 
to the 1D instantaneous reionization curve. Both approaches produce 
consistent probability distributions; the former method is used for 
$P(\tau)$ plotted in Figure~\ref{fig:tauinst}.

\begin{figure}
\centerline{\psfig{file=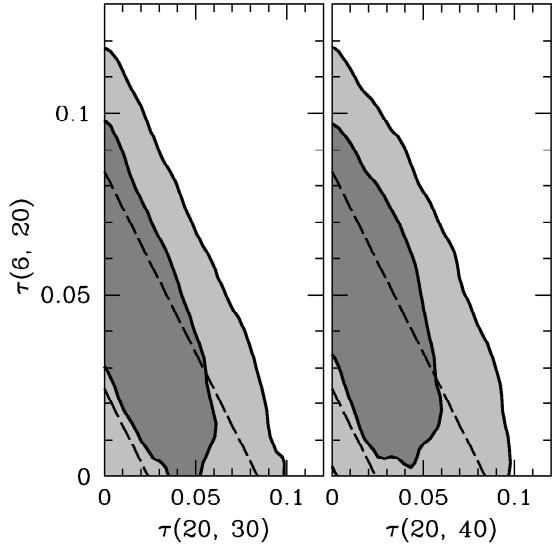, width=3.0in}}
\caption{Constraints on optical depth during reionization divided between 
two wide ranges in redshift, $\tau(6,20)$ and $\tau(20,\zmax)$, for 
$\zmax=30$ (\emph{left}) and 40 (\emph{right}). 
The contours are drawn at 68\% (\emph{dark gray}) and 95\% (\emph{light gray}) 
2D marginalized posterior probability. 
Dashed lines in each panel are lines of constant total optical depth: 
$\tau=0.12$ for the upper line, and $\tau=0.06$ for the lower line. 
\vskip 0.25cm
}
\label{fig:tauwmap}
\end{figure}

The weak $m_2$ constraint visible in Figure~\ref{fig:mcmcwmap} suggests 
that in addition to determining the total optical depth, 
\wmap\ data may provide useful limits on high-$z$ and low-$z$ 
optical depth as defined in the previous section. 
The 68 and 95\% posterior probability contours for the 
partial optical depths from \wmap\ 
are shown in Figure~\ref{fig:tauwmap} for 
$\zmid = 20$ and $\zmax = \{30,40\}$. 
The $x_e=0$ physicality prior cuts off the contours at $\tau(z_1,z_2)=0$; 
in both panels, the upper limits set by $x_e=1$ are outside the plotted area, 
so the contours are not strongly influenced by those priors.
The error ``ellipses'' are narrowest in the 
direction along which the total optical depth is constrained, as shown 
by dashed lines of constant total optical depth at $\tau=0.06$ and 
$\tau=0.12$ [approximately the upper and lower 1 $\sigma$ limits from 
the 3-year \wmap\ analysis of \citet{Speetal06}].

% ****************************************
% ****************************************

Comparison of the two panels in Figure~\ref{fig:tauwmap} 
reveals that the choice of $\zmax$ 
does not have a large effect on constraints on the
 optical depth above and below $z=20$.  This suggests
that $\tau(20,\zmax)$ should be interpreted as $\tau(z>20)$.  
We provide further
justification for this interpretation in \S~\ref{sec:cvdata}.

The 95\% upper limits on the high-redshift optical depth, 
$\tau(z>20)$, 
are $0.076$ and $0.078$ for $\zmax=30$ and 40, respectively, 
after marginalizing over all other parameters.
This is not a particularly strong constraint, since 
this result is only marginally inconsistent with all of the optical depth from 
reionization coming from $z>20$, but with future data it should be 
possible to either reduce the upper limit on high-$z$ optical depth 
or to detect the presence of a substantial ionized fraction at high redshift 
(see \S~\ref{sec:cvdata}). 

Models of reionization can be tested by computing the 
principal components of 
proposed ionization histories and comparing with constraints on 
$\{m_{\mu}\}$ from Monte Carlo chains.  
The \wmap\ constraints in Figure~\ref{fig:mcmcwmap} 
 are non-Gaussian, in part 
because the physicality priors on $\{m_{\mu}\}$ intersect the 
posterior probability contours where the likelihood is large. 
Because the constraints do not have a simple Gaussian form, it appears 
that the full parameter chains are necessary for accurate model testing, 
although the viability of various models can be estimated by comparing 
their eigenmode amplitudes with marginalized constraints such as 
those in Figure~\ref{fig:mcmcwmap}, keeping in mind that models favored 
by marginalized constraints could be disfavored in the full $N$-D 
parameter space.

% --------------------------------------------
\subsection{Cosmic variance-limited data}
\label{sec:cvdata}

To forecast how well principal components of the reionization history could be 
measured by future CMB polarization experiments, we repeat the analysis 
of \S~\ref{sec:wmap} using simulated realizations of the full-sky, 
noiseless $E$-mode angular power spectrum instead of \wmap\ data. 
Any real experiment will of course involve sky cuts, 
noise, foregrounds, and other complications, so the results 
presented here represent an optimistic limit on what we can learn about 
the global reionization history 
from low-$\ell$ $E$-mode polarization. At the end of this section, we 
estimate how much the constraints might be degraded from this idealized 
case for an experiment with characteristics similar to those proposed 
for the \emph{Planck} satellite.

We generate simulated realizations of $\clee$ 
drawn from $\chi^2$ distributions with cosmic variance determined by 
the theoretical angular power spectra that we compute using CAMB. 
For the $j$th sample in a chain, the likelihood 
including only $E$-mode polarization is
\begin{equation}
-\ln L_{(j)}=\sum_{\ell}\left(\ell+\frac{1}{2}\right)
\left(\frac{\hat{C}_{\ell}^{EE}}
{C_{\ell(j)}^{EE}}+\ln\frac{C_{\ell(j)}^{EE}}
{\hat{C}_{\ell}^{EE}}-1\right),
\label{eq:like}
\end{equation}
where $\hat{C}_{\ell}^{EE}$ is the spectrum of the simulated data and
$C_{\ell(j)}^{EE}$ is the theoretical spectrum calculated with the 
parameter values at step $j$ in the chain.

We have run Monte Carlo chains for multiple realizations of 
$\clee$ drawn from spectra computed assuming a variety of ``true'' 
reionization histories, $\xet(z)$.
%
% The ability of the simulated data to 
%constrain the principal components tends to be similar regardless 
%of the choice of $\xet(z)$, so 
We start by taking  the 
true history to be a model with nearly instantaneous reionization and 
$\tau=0.105$.  As a contrasting model for comparison we also use 
an extended, double reionization model with $\tau=0.090$. 
Figure~\ref{fig:cl} shows $\clee$ and $x_e(z)$ for 
each of these models. 

\begin{figure*}
\centerline{\psfig{file=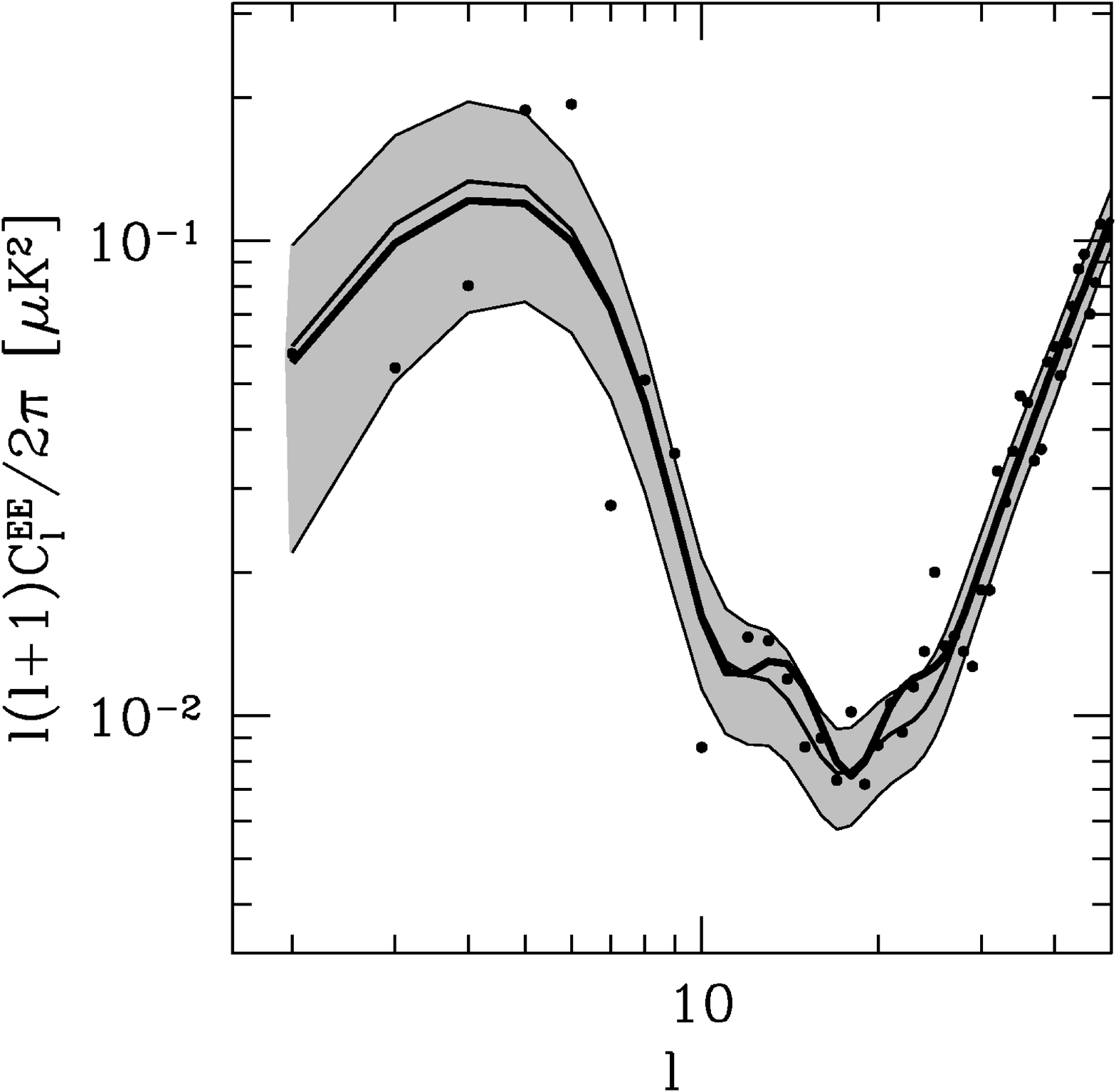, width=3.0in}
\psfig{file=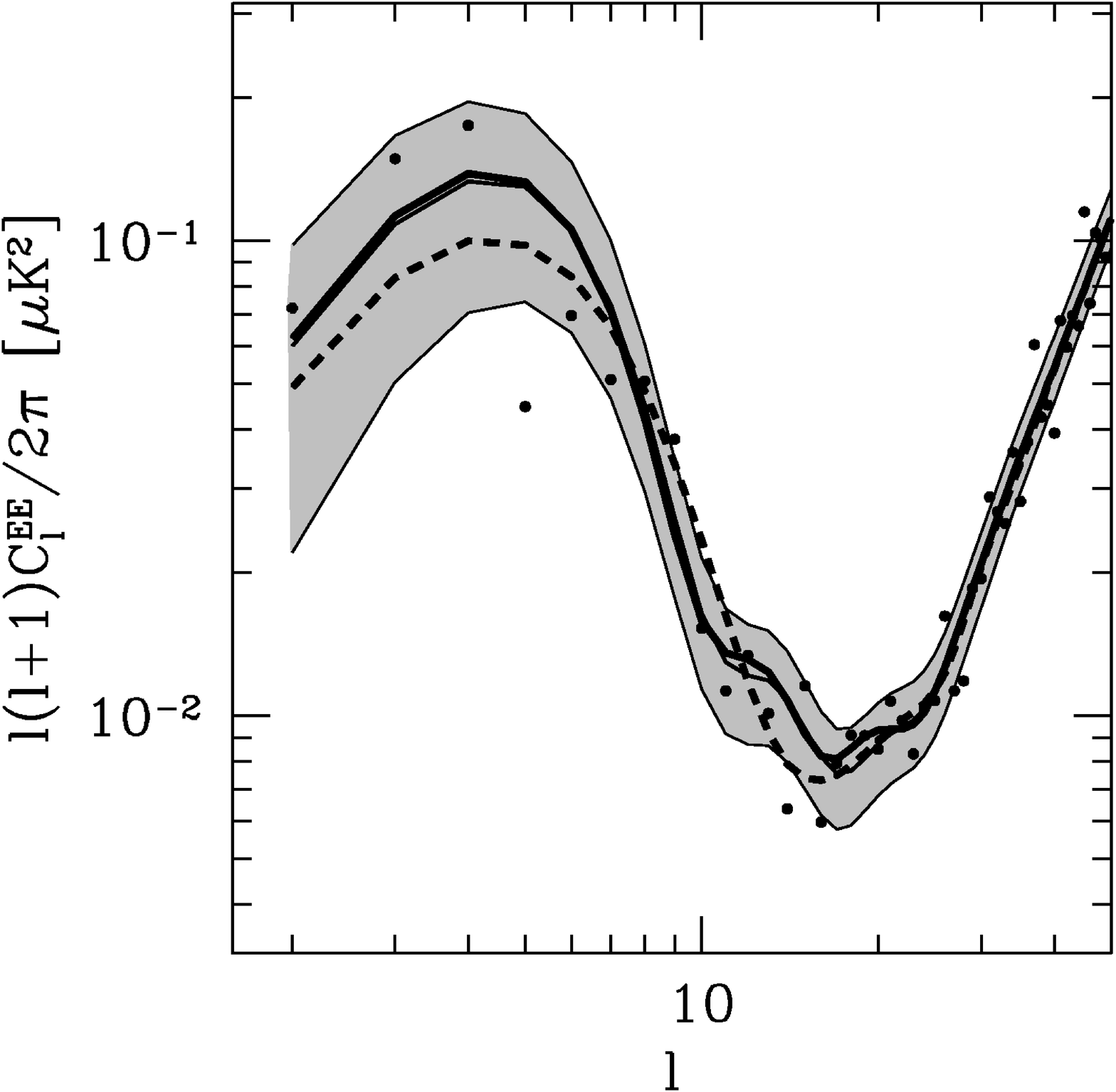, width=3.0in}}
\caption{$E$-mode spectra of the best-fit models from two Monte Carlo 
chains (\emph{thick curves}), each using a different realization of 
simulated cosmic variance-limited data (\emph{points}).  The 
realizations are drawn from the same $\clee$ of the instantaneous 
reionization model in Figure~\ref{fig:cl}, 
plotted as thin curves with shaded regions showing 
the $1~\sigma$ cosmic variance limits.
The dashed curve in the right panel shows the best fit model within the 
high-$m_2$ peak in the posterior probability (see Fig.~\ref{fig:mcmccv}).
\vskip 0.25cm}
\label{fig:clcv}
\end{figure*}

Since parameter constraints derived using 
a single draw of $\clee$ from the underlying power spectrum 
may contain features unique to that realization, we generated Monte Carlo 
chains for 10 random realizations of  
the instantaneous reionization 
power spectrum. Two of these realizations are 
plotted as points in Figure~\ref{fig:clcv}, along with the 
theoretical spectrum with cosmic variance bands. The thick curves 
in Figure~\ref{fig:clcv} are spectra of the best-fit models 
from the Monte Carlo chains for each realization. 
The overall best-fit models are similar for 
the two realizations and both agree closely with the theoretical $\clee$. 
The dashed curve in the right panel of Figure~\ref{fig:clcv} shows an 
``alternative'' best-fit model that we discuss later.

% ****************************************
% ****************************************

% ****************************************
\begin{figure*}
\centerline{\psfig{file=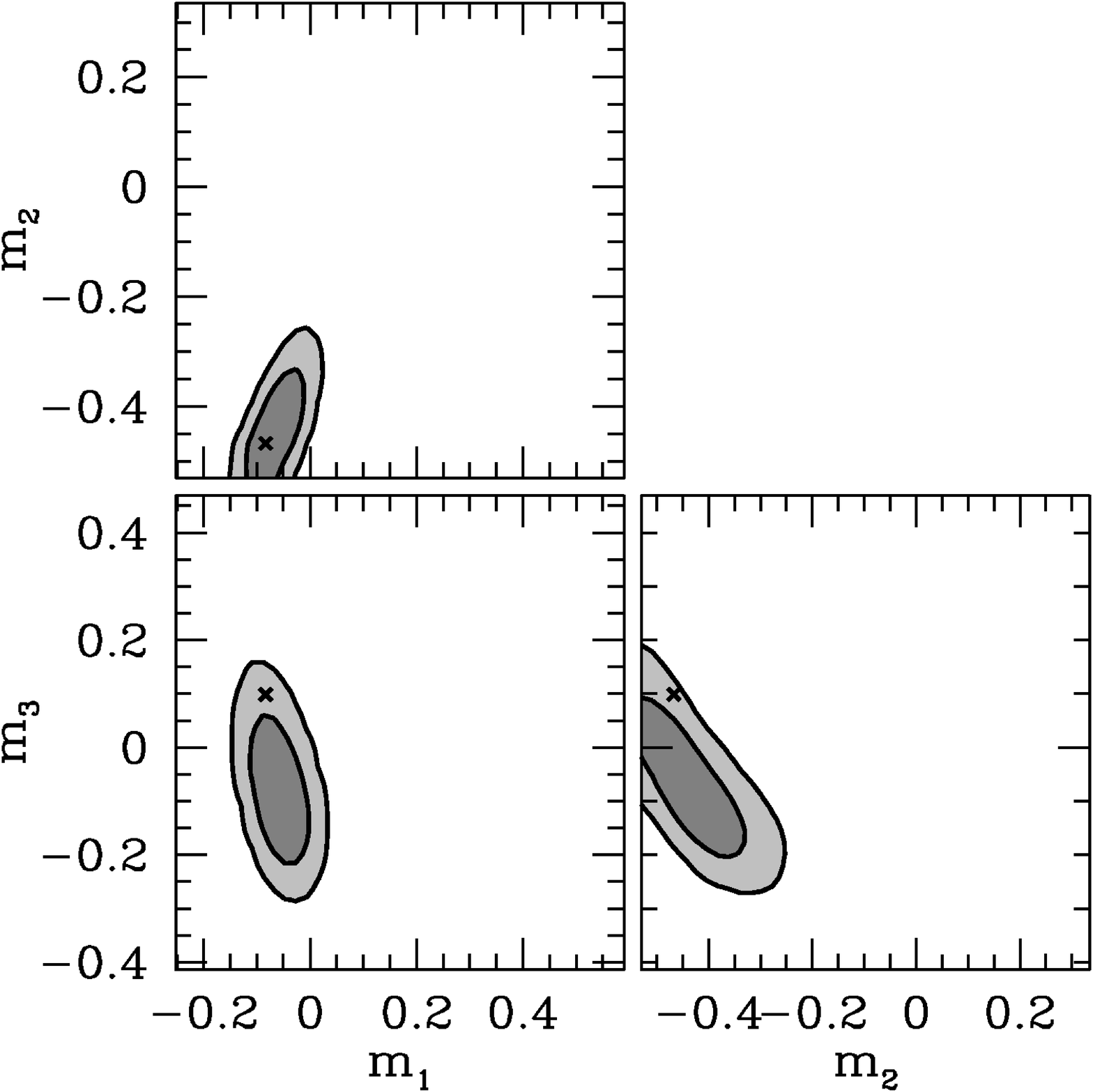, width=3.0in}
\psfig{file=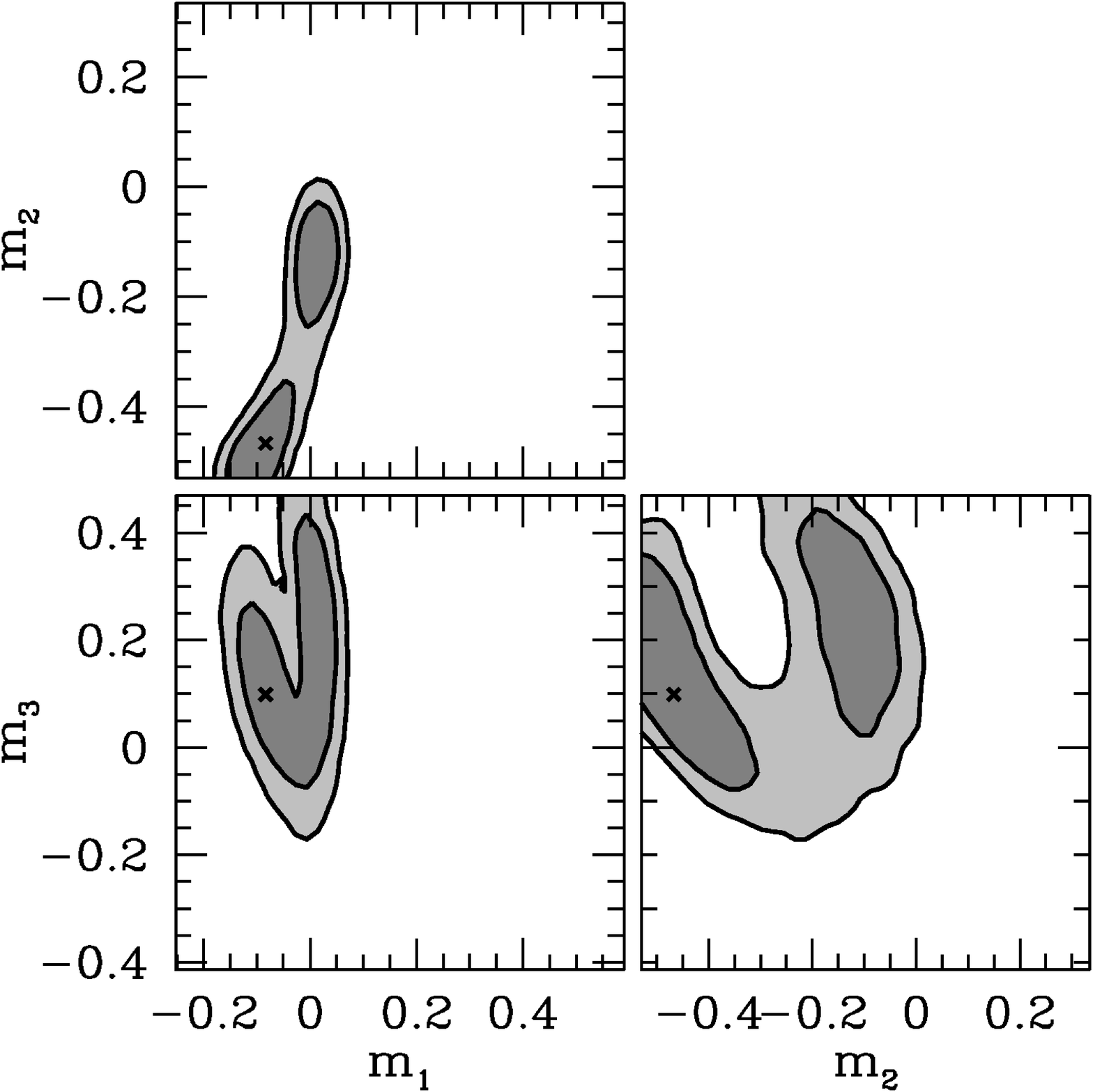, width=3.0in}}
\caption{Marginalized 2D contours (68 and 95\%) for 
the first three principal components with two different 
realizations of cosmic variance-limited data simulated using the 
instantaneous reionization history. The fiducial model has 
maximum redshift $\zmax=20$, and only the three modes shown here 
are varied in the parameter chains. The cross in each panel marks the 
location of the principal components of $\xet(z)$.
\vskip 0.25cm}
\label{fig:mcmccv}
\end{figure*}
% ****************************************

The 2D marginalized MCMC constraints on principal components for 
these two realizations are shown in Figure~\ref{fig:mcmccv}.
The plotted regions are bounded by the physical top-hat priors on $\{m_{\mu}\}$
as in Figure~\ref{fig:mcmcwmap}, but the eigenmodes are different here 
with $\zmax=20$ instead of 30 or 40 for the fiducial history. 
Since we have constructed $\xet(z)$ to have no ionization at $z\gtrsim 15$, 
we know that $\zmax=20$ should be a large enough maximum redshift; 
this choice of $\zmax$ also reflects the fact that improved empirical 
determination of $\zmax$ will allow it to be set at lower redshift 
for future data analysis, assuming that high-$z$ reionization is 
not detected.

The contours in Figure~\ref{fig:mcmccv} show that cosmic variance-limited 
data can constrain models within the physically allowed parameter space 
much better than current data can. The chains used for the results 
shown here vary the first 3 eigenmodes; by running chains with $N>3$ we 
find that cosmic variance-limited data can provide $2~\sigma$ constraints 
that are tighter than the physicality bounds for the first 4 eigenmodes 
of a $\zmax=20$ fiducial model and for at least 5 eigenmodes when $\zmax=30$.

Although the only difference in the MCMC setup for the left and right panels 
of Figure~\ref{fig:mcmccv} is the realization 
of $\clee$ (Fig.~\ref{fig:clcv}), 
there are some significant differences in the resulting 
eigenmode constraints, particularly in the $m_2-m_3$ plane. This kind of 
variation is typical among the realizations of simulated data, and 
is related to whether features in the theoretical $\clee$ remain intact 
with the inclusion of cosmic variance. The relevant features in the 
spectrum are the small oscillations in the low-power ``valley'' in 
$\clee$ on scales $10\lesssim \ell \lesssim 30$. In the example shown here, 
the feature that matters most is the bump in the theoretical spectrum at 
$\ell \approx 13$. In the left panel of Figure~\ref{fig:clcv}, the 
$\clee$ realization retains this bump, while in the right panel the 
bump is washed out by cosmic variance. The result is that the left 
realization is better able to pick out those reionization models 
in Monte Carlo chains that 
reproduce the true $\clee$, leading to the tight constraints on the left 
side of Figure~\ref{fig:mcmccv}. The realization in the right panel of 
Figure~\ref{fig:clcv} lacks this important ``fingerprint'' for identifying 
models that match the theoretical $\clee$, so the data allow a wider 
variety of models as reflected in the constraints on the right side of 
Figure~\ref{fig:mcmccv}.  In this case, there are two best-fit models 
corresponding to the two $68\%$ contours separated in the 
$m_2$ direction. The spectrum of the 
overall best-fit model (with smaller $m_2$), which is close to the 
theoretical spectrum, is 
plotted as a thick solid curve in the right panel of Figure~\ref{fig:clcv}, 
while the other best-fit model (at larger $m_2$) is plotted as a dashed 
curve. These two models have significantly different spectra, but 
because of the scatter in the random draw of $\clee$ they are each able 
to fit the data better at some multipoles and worse at others in such a way 
that the high-$m_2$ best-fit model is only a slightly worse fit than  
the low-$m_2$ best-fit model.

Among the realizations of simulated data that we explored with MCMC 
methods, constraints like those in the right panel of Figure~\ref{fig:mcmccv} 
are a fairly extreme case, occurring about 
10-20\% of the time. In general the realizations form a sort of 
continuum between the two presented here, with about half showing some 
displacement of the $95\%$ contour towards larger $m_2$ but 
not as much as in the second realization that we show as 
an example.

The extent to which differences between realizations show up in 
principal component constraints depends on $\xet(z)$. For example, 
the $\clee$ for the double reionization model (Fig.~\ref{fig:cl}) 
have more pronounced oscillations at $\ell\sim 10$-$30$ than the 
instantaneous reionization spectrum, so they are not as easily 
erased by cosmic variance or noise and the resulting principal component 
constraints tend to be more consistent from one realization to the next 
and more like the left panel of Figure~\ref{fig:mcmccv}.

The differences between constraints from various realizations of simulated 
data suggest the interesting possibility that our ability to learn about 
the reionization history from large scale $E$-mode polarization may 
ultimately depend on the luck of the draw of $\clee$ at our particular 
vantage point.  In an unlucky
 draw, cosmic variance can distort subtle features in the power 
spectrum in a way that would limit constraints on reionization 
eigenmodes to be worse than we might expect from the Fisher approximation 
or a more typical draw. The good news is that 
even a realization like the one in the right panels of Figures~\ref{fig:clcv}
and~\ref{fig:mcmccv} would permit constraints that are far better than 
what is currently possible.  It is also important to note that although such 
constraints are weaker than in the best-case scenario, 
they are still consistent with the true 
parameter values and in general would not lead us to rule out the 
true reionization history based on a principal component analysis 
of the data.

As with the principal components of $x_e(z)$, MCMC analysis indicates that 
constraints on the total optical depth to reionization would be  
greatly improved with cosmic variance-limited data: 
$\sigma_{\tau}\approx 0.005$, down a factor of six from the 3-year 
\wmap\ value, $\sigma_{\tau}\approx 0.03$. 
The constraints on 
$\tau$ using the realization of instantaneous reionization $\clee$ 
in the left panel of Figure~\ref{fig:clcv} are listed in Table~\ref{tab:tau} 
for two fiducial models with $\zmax=20$ and 30. 
Optical depth constraints from the other draw of $\clee$ are similar, 
so estimates of $\tau$ and $\sigma_{\tau}$ appear not to be biased 
for atypical realizations. It makes sense that differences in realizations 
do not significantly affect $\tau$ since the optical depth constraint 
comes mainly from the peak in $\clee$ on the largest scales whereas 
differences in principal component constraints between realizations 
arise from the low-power part of the spectrum at higher $\ell$.

The model-independent constraints on $\tau$ are neither biased nor 
significantly weaker relative to  
the constraint obtained assuming instantaneous reionization 
(bottom row in Table~\ref{tab:tau}), as is true for the \wmap\ data. 
This should not be surprising since we have assumed that $\xet(z)$ is 
an instantaneous reionization model. 
However, if $\xet(z)$ is in fact very different from the instantaneous 
model, the estimate of $\tau$ obtained under the assumption of 
instantaneous reionization will be incorrect. 
For example, if we take $\xet(z)$ to be the double reionization 
model of Figure~\ref{fig:cl}, the model-independent approach yields 
an optical depth constraint consistent with the true value of 
$\tau=0.090$. On the other hand, MCMC analysis restricted to 
instantaneous reionization histories gives a significantly biased 
estimate, $\tau=0.135\pm 0.005$. In this case, the 1D curve of instantaneous 
reionization models in the space of eigenmodes never 
lies near the principal component values favored by the data, so any 
analysis that only considers such models would never find a 
good fit to the data.  This illustrates the importance of a model-independent 
approach: although results from current data may not be significantly 
affected by the assumption of a specific form of $x_e(z)$, future 
data will be sensitive enough to the reionization history that 
the choice of a specific model can greatly 
impact the results \citep{Holetal03,Coletal05}.

% ****************************************
\begin{figure*}
\centerline{\psfig{file=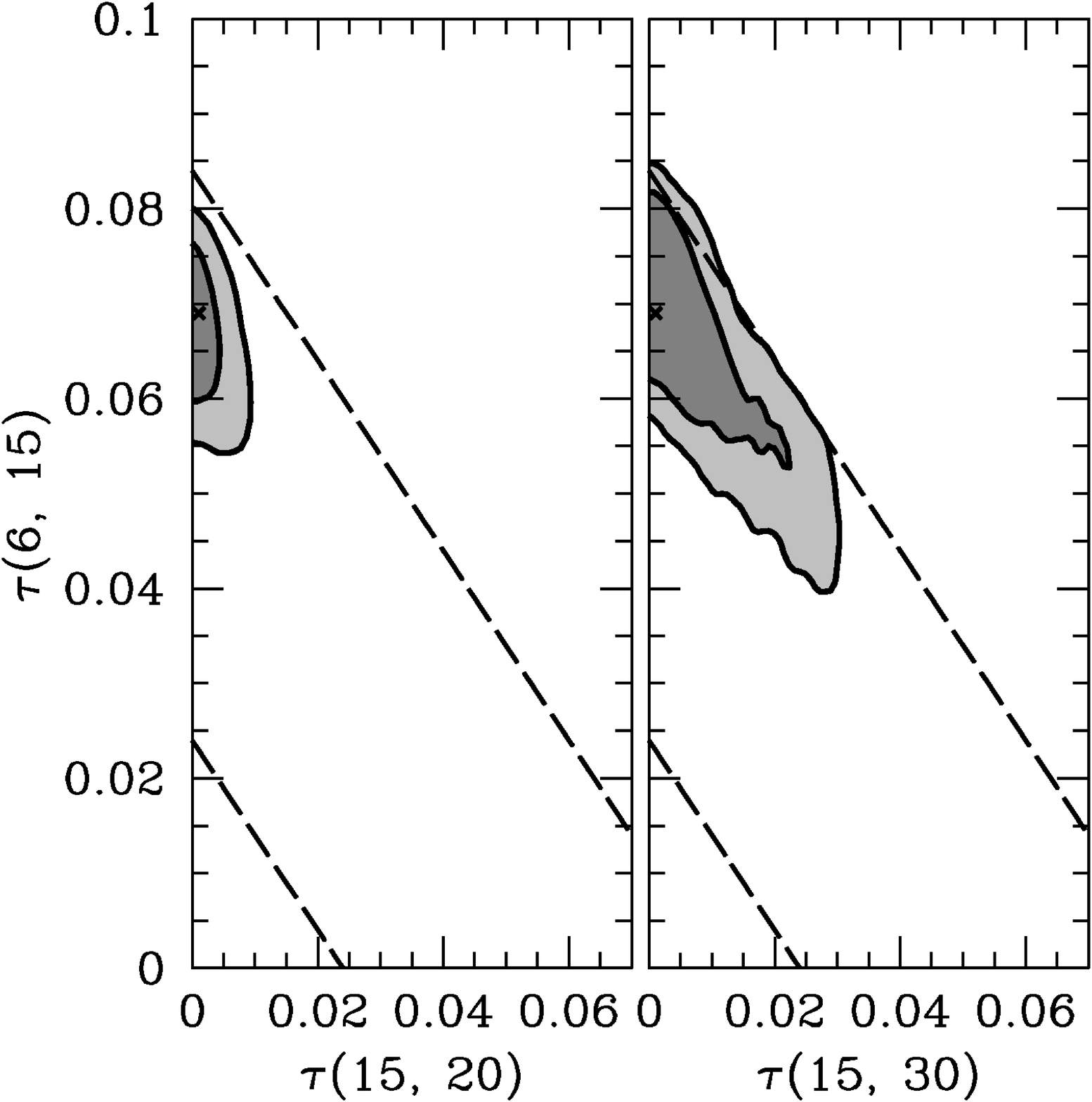, width=3.0in}
\psfig{file=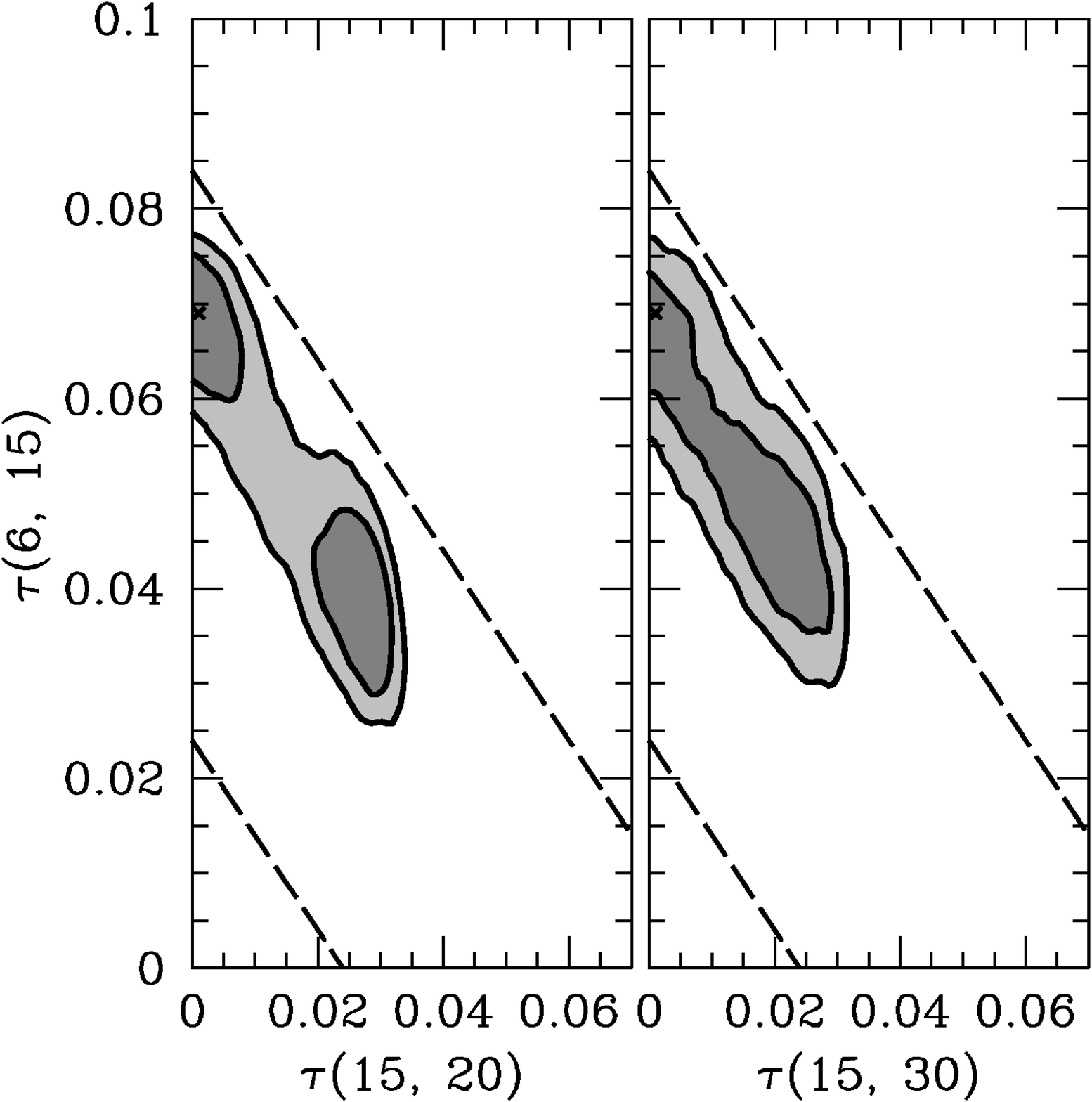, width=3.0in}}
\caption{Same as Figure~\ref{fig:tauwmap} for simulated 
cosmic variance-limited data using the two realizations 
of instantaneous reionization $\clee$ shown in 
Figure~\ref{fig:clcv}. Crosses indicate the 
partial optical depths for the true reionization history. 
The redshifts chosen for $\zmid$ and $\zmax$ are also lower here than  
for the \wmap\ analysis, reflecting the expectation that 
future data may permit better 
empirical constraints on $\zmax$.
\vskip 0.25cm}
\label{fig:taucv1}
\end{figure*}
% ****************************************

% ****************************************
\begin{figure}
\centerline{\psfig{file=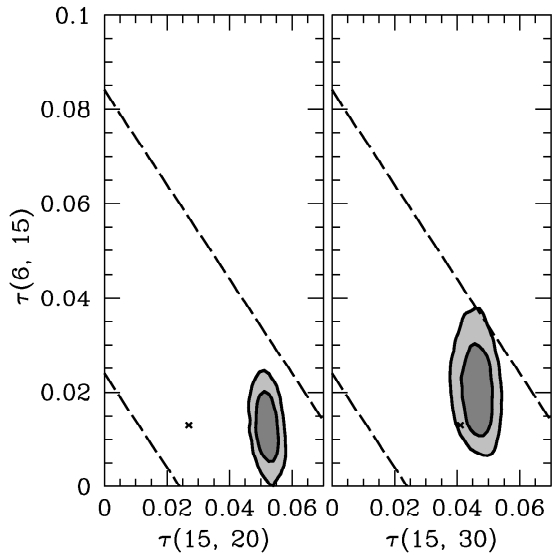, width=3.0in}}
\caption{Same as Figure~\ref{fig:taucv1}, but 
assuming that $\xet(z)$ is a double reionization history 
with $\zmax=23$ (thin curves in Figure~\ref{fig:cl}). 
Crosses mark the true partial optical depths for this model; 
note that unlike in Figure~\ref{fig:taucv1}, $\tau(15,20)\neq \tau(15,30)$ 
for the expected values here since $\xet(z)$ has a nonzero 
ionized fraction between redshifts 20 and 30.
\vskip 0.25cm}
\label{fig:taucv2}
\end{figure}
% ****************************************

Though constraints on the contributions to $\tau$ from high and low $z$ 
are currently limited to weak upper bounds, 
future polarization data will likely enable more definitive detections 
of high-$z$ reionization if it is present, or tighter upper limits if absent.
The left and right panels of Figure~\ref{fig:taucv1} 
show these constraints from Monte Carlo chains using the 
$\clee$ plotted in the left and right panels of Figure~\ref{fig:clcv}, 
respectively. The true ionization history has zero ionized fraction at 
$z>15$, and the MCMC analysis places a 95\% CL upper limit 
on $\tau(z>15)$ of $\sim 0.03$.
Figure~\ref{fig:taucv2} shows MCMC constraints on the same partial 
optical depths using simulated data where the true model is instead 
taken to be the extended, double reionization history plotted as 
a thin curve in the inset of Figure~\ref{fig:cl}. In this case, most 
of $\tau(\zmin,\zmax)$ comes from $z>15$, and using the principal 
components of $x_e(z)$ the high-$z$ optical depth can be detected at 
a high level of confidence and measured to an accuracy of $\sim 0.01$.
Note that the constraints in both Figures~\ref{fig:taucv1} 
and~\ref{fig:taucv2} are consistent with the \wmap\ constraints 
in Figure~\ref{fig:tauwmap}.

As with \wmap\ data, most of the 
constraints from simulated data in Figures~\ref{fig:taucv1} 
and~\ref{fig:taucv2} do not appear to depend strongly on the 
choice of $\zmax$. For the instantaneous reionization model used for 
Figure~\ref{fig:taucv1}, this is to be expected since we know that the 
simulated $\clee$ have no contribution from $z \gtrsim 20$. 
Each set of contours in Figure~\ref{fig:taucv1} is consistent with 
the values of $\tau(6,15)$ and $\tau(15,\zmax)$ for $\xet(z)$. 
The constraints in the second panel in Figure~\ref{fig:taucv1} 
extend to larger high-$z$ optical depth than those in the first panel 
because the effect of the physicality priors is not as strong at 
larger $\zmax$, as noted in \S~\ref{sec:wmap}; the leftmost 
$\zmax=20$ contours would stretch to larger $\tau(15,20)$ and 
smaller $\tau(6,15)$ if we did not apply the physicality bounds.
The two panels on the right side of Figure~\ref{fig:taucv1} show 
weaker constraints on the partial optical depths because 
the principal component constraints for that realization of $\clee$ 
are weaker.

In the extended, double reionization model used for 
Figure~\ref{fig:taucv2}, reionization 
actually starts at $z=23$, so the choice of $\zmax=20$ is not 
appropriate for this model. The result of such an error is that 
the constraint on $\tau(15,20)$ is inconsistent with the true 
value, marked by a cross in the left panel of Figure~\ref{fig:taucv2}. 
The $\zmax=30$ constraints in the right panel are consistent with 
the expected value, although not in perfect agreement due to 
cosmic variance. It is interesting that although the value chosen for 
$\zmax$ in the left panel is too low for this model, the 
constraints on $\tau(z>15)$ for $\zmax=20$ and $\zmax=30$ are still consistent 
with each other.  
This supports the interpretation of constraints on $\tau(\zmid,\zmax)$
as $\tau(z>\zmid)$ in \S~\ref{sec:wmap}

While the cosmic variance-limited simulated data are useful for 
determining how well the reionization history could possibly be 
constrained by large-scale $E$-mode polarization measurements, it is 
also interesting to ask how well we can do with future experiments 
that fall somewhat short of the idealized case that we have considered 
so far. In particular, the upcoming \emph{Planck} satellite is 
expected to improve our knowledge of the large-scale $E$-mode spectrum 
substantially \citep{Planck}; what does this imply for constraints on $x_e(z)$?
To estimate what might be possible with \emph{Planck} data, 
we assume that after subtracting foregrounds 
a single foreground-free frequency channel remains 
for constraining the low-$\ell$ 
$E$-mode polarization. We take this to be the 143 GHz channel with 
a white noise power level of $w_P^{-1/2}=81~\mu$K$'$ and 
beam size $\theta_{\rm FWHM}=7.1'$, and we assume that the sky coverage is 
$f_{\rm sky}=0.8$ after cutting out the Galactic 
plane~\citep{Planck,Albetal06}. We compute the likelihood of Monte Carlo 
samples using the routines provided in CosmoMC 
and analyze parameter chains with the principal component method as 
described for \wmap\ and cosmic variance-limited data.

For various choices of $\xef(z)$ and $\xet(z)$ we find that going from 
full-sky cosmic variance-limited data to \emph{Planck}-like data 
increases the uncertainty in $x_e(z)$ principal components, $\sigma_{\mu}$,
and in the optical depth, $\sigma_{\tau}$, by roughly a factor of two.
Based on these results, it should be possible to constrain about  
three eigenmodes at $\sim 2~\sigma$ 
within the space of physical models, and we expect 
$\tau$ to be determined to an accuracy of about $\pm 0.01$.
This is consistent with previous estimates of the \emph{Planck} 
optical depth uncertainty when considering more general reionization 
models than instantaneous reionization \citep[e.g.,][]{Holetal03,HuHol03}.

% --------------------------------------------
\subsection{Fisher matrix predictions versus MCMC results}
\label{sec:fisher}

The Fisher matrix analysis of \S~\ref{sec:eigenmodes} predicts that 
the principal components of $x_e(z)$ should be uncorrelated, with 
errors given by $\sigma_{\mu}$ from equation~(\ref{eq:fisher2}) 
(in the idealized limit of full sky, noiseless observations). 
These characteristics rely on the assumption that the difference between 
the true and fiducial reionization histories is small: 
$\delta \xet(z)=\xet(z)-\xef(z) \ll 1$. Clearly this assumption will 
not hold for any single fiducial history if one wants to consider 
a variety of possible true histories, and the consequence is that 
$\langle m_{\mu}m_{\nu} \rangle$ differs from $\sigma_{\mu}^2 \delta_{\mu\nu}$.

% ****************************************
\begin{figure}
\centerline{\psfig{file=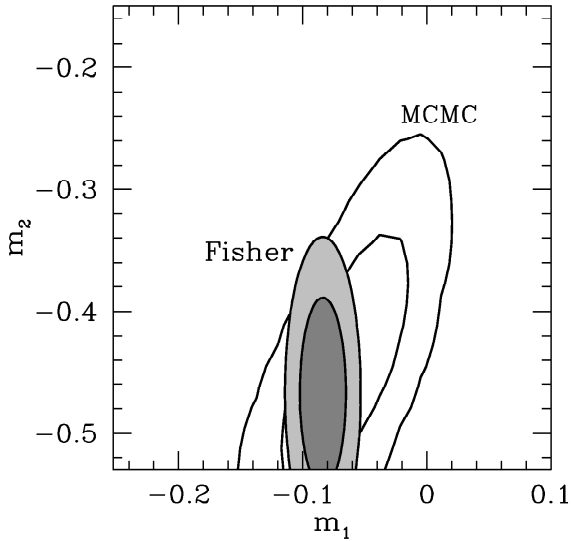, width=3.0in}}
\caption{Monte Carlo constraints in the $m_1$-$m_2$ plane from the 
left panel of Figure~\ref{fig:mcmccv} (68 and 95\% contours) compared 
with the estimated constraints from the Fisher matrix 
(\emph{shaded contours}).
\vskip 0.25cm
}
\label{fig:fisher2d}
\end{figure}
% ****************************************

As an example of the difference between the Fisher approximation and 
the full Monte Carlo analysis, in Figure~\ref{fig:fisher2d} 
the marginalized constraints in the $m_1-m_2$ plane from the 
left panel of Figure~\ref{fig:mcmccv} 
are compared with the Fisher matrix error ellipses, centered on 
the true $(m_1,m_2)$ 
for the same choices of $\xet(z)$ (instantaneous reionization 
with $\tau=0.105$) and $\xef(z)$ (constant $x_e=0.15$ out to $\zmax=20$). 
The MCMC constraints have correlated values of $m_1$ and $m_2$ with
somewhat larger uncertainties than the Fisher matrix $\{\sigma_{\mu}\}$. 
The eigenfunctions, $S_{\mu}(z)$, are constructed to have orthogonal 
effects on $\clee$ in 
the vicinity of the fiducial model, but for large $\delta x_e(z)$  
the orthogonality breaks down.  Fortunately the errors on $\tau$ 
remain largely unaffected because the
$m_{1}-m_{2}$ correlation is oriented 
such that the increased errors
are along the direction of constant total $\tau$ (see Figure~\ref{fig:tauinst}).

One can compute new eigenfunctions 
by evaluating ${\partial \ln \clee}/{\partial x_e(z)}$ at $\xet(z)$ 
and see that their dependence on redshift 
differs from the original $S_{\mu}(z)$. 
It is possible to decorrelate the principal components by 
diagonalizing the covariance matrix from the Monte Carlo chains 
and rotating the eigenmodes into the new basis. 
However, this procedure is unnecessary for most applications; 
the most important property of the principal components for 
constraining $x_e(z)$ is that the first few modes form a complete basis 
for $\clee$ on large scales, accurate within cosmic variance limits. 
Completeness of the eigenmodes holds true even if they are not
exactly orthogonal.

Finally, note that if the true ionization history and the draw of $\clee$ 
permit constraints on principal components similar to those in the 
left panel of Figure~\ref{fig:mcmccv}, then these constraints may be 
close enough to Gaussian so that the covariance matrix would be 
sufficient to describe the information in the $\clee$ for 
finding constraints on reionization models. However, if what we see looks 
more like the right panel of Figure~\ref{fig:mcmccv} then the full 
chains of Monte Carlo samples may be necessary for further applications 
even if we can measure the $E$-mode polarization perfectly.

% =====================================================
\section{Discussion}
\label{sec:discuss}

Observations of the large-scale $E$-mode polarization of the CMB in 
the near future are expected to yield new information 
about the spatially-averaged reionization history of the universe. 
The principal components of the reionization history are a 
promising tool for extracting as much of that information as possible 
from the data. We have shown that the principal component method 
can be usefully applied to real, currently available data, and 
forecasts from simulated data suggest that there is room to 
substantially improve constraints on the reionization history 
using this method as measurements of the large-scale $E$-modes improve.

We find that the key features of the principal component analysis 
put forward by \cite{HuHol03} continue to apply when we go from 
the Fisher matrix approximation to an exploration of the full 
likelihood surface using Markov Chain Monte Carlo methods. 
For fairly conservative choices of the maximum redshift of reionization 
($\zmax \sim 30$-40), only the first five principal components at most 
are needed for a complete representation of the $E$-mode angular 
power spectrum to within cosmic variance. To account for 
arbitrary reionization histories in the analysis of CMB data, only a 
few additional parameters must be included in chains of Monte Carlo 
samples if those parameters are taken to be the lowest-variance 
principal components of $x_e(z)$. 

Specific models of reionization can be tested easily by computing 
their eigenmode amplitudes and comparing with constraints on the 
eigenmodes from the data. Constraints on derived parameters, 
such as total optical depth or the optical depth from a certain 
range in redshift, represent other applications of the 
MCMC constraints on principal components of $x_e(z)$.

Often, estimates of the optical depth to reionization 
are computed assuming instantaneous reionization or some other simple 
form for $x_e(z)$. Here we extend the analysis of the 3-year \wmap\ 
data to allow a more general set of models of the global 
reionization history. We find that expanding the model space does not 
significantly widen the uncertainty in $\tau$ beyond the 
instantaneous reionization value of $\sigma_{\tau} = 0.03$. 
Robust $\tau$ constraints are important for tests of the dark energy
based on the growth of structure since they control the uncertainty
on the amplitude of the initial spectrum.

Moreover, even with current data the principal component constraints 
are beginning to show the possibility of determining properties of 
reionization in addition to $\tau$. By comparing the optical depth from 
high $z$ with that from low $z$, we obtain an upper limit on the 
contribution to the optical depth from high redshift: 
$\tau(z>20) < 0.08$ at 95\% confidence, assuming that there is no 
significant episode of reionization at $z\gg 40$.

Due to the limitations of noise and foreground contamination, 
only the first two eigenmodes of the reionization 
history, $m_1$ and $m_2$, can be determined with present 
polarization data to any reasonable degree of 
accuracy. Constraints on these two modes come primarily from the 
main, broad peak in $E$-mode power at low $\ell$ from reionization. 

As measurements of the $E$-mode polarization improve, for example from 
additional \wmap\ data or through planned future experiments such as 
\emph{Planck}, better knowledge of the low-power ``trough'' in $\clee$ 
between the main reionization peak and the first acoustic peak 
should enable constraints on the third and higher principal components, 
up to about $m_5$ for near cosmic variance-limited data. 
Since constraints on these higher modes rely on the ability to identify 
subtle features in the trough of $\clee$, the ultimate accuracy to 
which the eigenmodes can be determined may depend on whether or not 
the necessary features are well reproduced in the particular 
random draw of $\clee$ that is available to us. However, even if we 
are unlucky enough to have a realization in which some of the important 
features of the spectrum are washed out by randomness, it should still 
be possible to measure several of the principal components to 
better accuracy than is currently possible. Knowledge of the $\mu\geq 3$ 
eigenmodes of $x_e(z)$, along with improved 
constraints on $m_1$ and $m_2$, will 
allow more stringent tests of reionization models and a better 
understanding of the global reionization history.

\acknowledgements {\it Acknowledgments}:  We thank Cora Dvorkin,
Gilbert Holder, Dragan Huterer, Hiranya Peiris, and Jochen Weller
for useful discussions.  MJM was supported by a 
National Science Foundation Graduate Research Fellowship. 
WH was supported by the KICP
through the grant NSF PHY-0114422, the DOE through 
contract DE-FG02-90ER-40560 and the David and
Lucile Packard Foundation.

\bibliographystyle{apj}
%\bibliography{ms}

\end{document}